\documentclass{aa}
\pdfoutput=1

\usepackage[varg]{txfonts}

\usepackage{multirow}
\usepackage{natbib}
\bibpunct{(}{)}{;}{a}{}{,} 

\begin{document}
\title{The evolution of the jet from Herbig Ae star HD 163296 from
  1999 to 2011\thanks{Figures 1-2 are only available in electronic form via
http://www.edpsciences.org}}

\author{H.~M.~G\"unther \and P.~C.~Schneider \and Z.-Y.~Li}
\institute{Harvard-Smithsonian Center for Astrophysics, 60 Garden Street, Cambridge, MA 02138, USA, hguenther@cfa.harvard.edu  \and Universit\"at Hamburg, Hamburger Sternwarte, Gojenbergsweg 112, 21029 Hamburg, Germany \and Department of Astronomy, University of Virginia, P.O. Box 400325, Charlottesville, VA 22904, USA}

\date{Received <date> Accepted <date>}

\abstract{Young A and B stars, the so-called Herbig Ae/Be stars (HAeBe), are surrounded by an active accretion disk and drive outflows. We study the jet HH~409, which is launched from the HAeBe star HD~163296, using new and archival observations from \emph{Chandra} and \emph{HST}/STIS. In X-rays we can show that the central source is not significantly extended. The approaching jet, but not the counter-jet, is detected in Ly$\alpha$.  In addition, there is red-shifted Ly$\alpha$ emission extended in the same direction as the jet, that is also absent in the counter-jet. We can rule out an accretion or disk-wind origin for this feature. In the optical we find the knots B and B2 in the counter-jet. Knot B has been observed previously, so we can derive its proper motion of $(0.37\pm0.01)$\arcsec{}~yr$^{-1}$. Its electron density is 3000~cm$^{-3}$, thus the cooling time scale is a few months only, so the knot needs to be reheated continuously. The shock speed derived from models of H$\alpha$ and forbidden emission lines (FELs) decreased from 50~km~s$^{-1}$ in 1999 to 30~km~s$^{-1}$ in 2011 because the shock front loses energy as it travels along the jet. Knot B2 is observed at a similar position in 2011 as knot B was in 1999, but shows a lower ionization fraction and higher mass loss rate, proving variations in the jet launching conditions.}


\maketitle

\section{Introduction}
Stars form when molecular clouds fragment and contract to proto-stars. Mass accretion onto those stellar cores proceeds via an accretion disk, while the surrounding envelope disperses. Eventually, the stars become optically visible; at this stage the low-mass population is called classical T~Tauri stars (CTTS), the A and B star progenitors are Herbig Ae/Be stars (HAeBe). The angular momentum accreted from the disk can be removed by outflows, which come in different types: There are slow, wide-angle winds, often seen in molecular lines such as H$_2$, faster winds, showing up in optical forbidden emission lines (FELs) such as [\ion{O}{i}], and highly collimated jets, which often reach velocities up to 400~km~s$^{-1}$ \citep{1998AJ....115.1554E}. In very young objects these jets can be traced out to a few parsecs and are still well collimated. Internal working surfaces and bow shocks are visible in emission lines as knots, the so-called Herbig-Haro (HH) objects. The evidence for X-rays from outflows is relatively recent, starting with \object{HH 2} \citep{2001Natur.413..708P,2012A&A...542A.123S} and \object{HH 154} \citep{2003ApJ...584..843B,2006A&A...450L..17F,2011A&A...530A.123S}. In both cases the jet driving sources are deeply embedded. 

To date, we know four CTTS or HAeBe with spatially extended X-ray emission: The CTTS \object{DG Tau} \citep{2005ApJ...626L..53G,2008A&A...478..797G,2011arXiv1101.2780G,Schneider,dgtau} and RY~Tau \citep{2011ApJ...737...19S}, Z~CMa, a young star in an FUOr outburst phase \citep{2009A&A...499..529S} and the HAeBe \object{HD 163296} \citep{2005ApJ...628..811S}. 

Jets from CTTS consist of layers with different velocities, where the innermost and most collimated component is the fastest one \citep{2000ApJ...537L..49B} with velocities of a few hundred km~s$^{-1}$. The outer layers are slower and less collimated. Often there is an asymmetry between the jet and the counter jet. 

Little is known about the physical mechanism, which drives the outflows. Different theoretical models of stellar winds \citep{1988ApJ...332L..41K,2005ApJ...632L.135M}, X-winds \citep{1994ApJ...429..781S} and disk winds \citep{1982MNRAS.199..883B,2005ApJ...630..945A} have been proposed. 
Ultimately, the jet launching must be powered from the gravitational energy released in the accretion process. This is supported by the observation that the outflow rate is roughly one tenth of the accretion rate \citep{1990ApJ...354..687C,1995ApJ...452..736H,2008ApJ...689.1112C}, but it is unclear how the energy is converted. 

All models of jet launching  and collimation rely on magnetic fields of some kind. However, jets are observed not only from solar-mass CTTS, but also from HAeBes. In contrast to CTTS with magnetic fields typically in the kG range, HAeBes are not expected to posses outer convection zones; thus, they cannot drive a solar-like dynamo, although primordial magnetic fields might be present. Often no or only weak fields can be observed \citep{2007A&A...463.1039H,2007MNRAS.376.1145W}; the formal limit for HD~163296 is $-25\pm 27$~G \citep{2007A&A...463.1039H}.

In this article we present new results on the time-evolution of the jet from the HAeBe HD~163296. 

In Sect.~\ref{sect:target} we summarize previous work done on HD~163296 and its jet HH~409. In Sect.~\ref{sect:obs} we give details on observations and data reduction. The results from the individual observations are given in Sect.~\ref{sect:results}, followed by a discussion (Sect.~\ref{sect:discussion}). We end with a summary in Sect.~\ref{sect:summary}.

\section{HD 163296 and its jet HH 409}
\label{sect:target}
HD~163296 is surrounded by an accretion disk with a mass accretion rate of $\dot
M_{acc}\approx10^{-7}$~M$_{\odot}$~yr$^{-1}$. This disk shows a
position angle of $(140 \pm 5)^{\circ}$ and an inclinaton of 
$i=(60\pm5)^{\circ}$ \citep{2006A&A...459..837G}\footnote{All position
angles in this work are measured North to East.}. Perpendicular to it
lies a chain of nebulosities \citep[position angle $(42.5\pm3.5)^{\circ}$][]{2000ApJ...544..895G}, called \object{HH 409}. This jet is detected both by coronagraphic imaging \citep{2000ApJ...544..895G,2006ApJ...650..985W} and in \emph{HST}/STIS long slit spectra \citep{2000ApJ...542L.115D}.

The distance to HD~163296 is $119^{+12}_{-10}$~pc
\citep{2007A&A...474..653V}. The presence of a late-type companion
can be ruled-out with a high degree of confidence from the \emph{HST}
imaging, the UV and optical spectra \citep[][and references
therein]{2005ApJ...628..811S} and also more recent high-resolution
optical spectroscopy is fully consistent with HD~163296 being single \citep{2009A&A...495..901M}.

The jet can be traced down to 0\farcs06 (7.1~AU) in the \emph{HST}/STIS image and out to $\approx 4000$~AU in narrow band optical imaging \citep{2006ApJ...650..985W}. Jet and counter jet are asymmetric. They have a different number of knots and different opening angles, but surprisingly the mass loss rate agrees better than a factor of 2 on both sides ($\dot M_{jet} \approx 10^{-8}$~M$_{\odot}$~yr$^{-1}\approx0.1\times\dot M_{acc}$). The opening angle is about $2^\circ$ for the jet; the initial value for the counter jet is $5^\circ$, but it is more collimated further out \citep{2006ApJ...650..985W}. 
The total space velocities, calculated from proper motion and radial velocity, are 360~km~s$^{-1}$ and 260~km~s$^{-1}$, respectively. The knots of the jets radiate in Ly$\alpha$, H$\alpha$ and optical FELs, e.g. [\ion{O}{i}], [\ion{S}{ii}] and [\ion{N}{ii}]. 
Line ratios of the last three ions are density and temperature sensitive and thus give estimates for the electron densities in the HH objects of the order $10^3$~cm$^{-3}$. \citet{2006ApJ...650..985W} show structural differences between the knots: In knot C the H$\alpha$ and [\ion{S}{ii}] emission are coincident, but in knot~A H$\alpha$ precedes the [\ion{S}{ii}] emission. We use the naming convention of \citet{2006ApJ...650..985W} throughout this work to refer to individual knots. 

HH~409 seems similar to jets from CTTS in mass loss rate, electron excitation and opening angle.
However, these quantities are all measured from a jet component which moves at an intermediate velocity of 200-300~km~s$^{-1}$. 
There are two indications that the fastest component of HH~409 is more energetic. Both come from X-ray observations: First, \citet{2005ApJ...628..811S} detected X-ray emission (luminosity  $3\times10^{27}$~erg~s$^{-1}$) close to knot A at 7~\arcsec (=830~AU) from the star. Second, the X-ray spectrum of the star itself shows the UV-sensitive \ion{O}{vii} triplet, whose line ratios are not compatible with emission close to the stellar surface. This suggests X-ray emission from jet collimation shocks (within a few AU of the star). Using the observed X-ray temperature,  \citet{HD163296}  estimate a flow velocity of 500~km~s$^{-1}$ for the fastest jet component. Yet, the optical FELs and the Ly$\alpha$/H$\alpha$ ratio indicate shock speeds of the order 80-90~km~s$^{-1}$ \citep{2006ApJ...650..985W}, which is far too slow to produce any X-ray emission.

\section{Observations and data reduction}
\label{sect:obs}

\begin{table*}
\caption{Log of observations\label{tab:obslog}}
\centering
\begin{tabular}{lllrlrrr}
\hline\hline
Facility & Instrument & Id & exposure & date & P.A. & central wavelength & aperture\\
\hline
Chandra & ACIS-S & 3733 & 20 ks & 2003-08-10\\
Chandra & ACIS-S & 12359 & 46 ks & 2011-02-09\\
HST & STIS/G750L & O5FO22010 & 1140 s & 1999-08-09 & 43.06\degr & 7751 \AA{} & 52X0.2F1\\
HST & STIS/G140M & O57Z03010 & 2230 s & 1999-09-30 & 45.15\degr & 1218 \AA{} & 52X0.2\\
HST & STIS/G140M & O66Q02010 & 2282 s & 2000-07-22 & 46.25\degr  & 1218 \AA{} & 52X0.5\\
HST & STIS/G140M & O66Q03010 & 7707 s & 2000-07-21 & 46.50\degr & 1218 \AA{} & 52X0.2\\
HST & STIS/G750M & OBI801010 & 2275 s & 2011-07-21 & 44.06\degr & 6581 \AA{} & 52X0.2F1\\
HST & STIS/G140M &  OBI801020 & 2700 s & 2011-07-21 & 44.00\degr & 1218 \AA{} & 52X0.2F1\\
\hline
\end{tabular}
\end{table*}
Table~\ref{tab:obslog} gives a summary of the observations we used. 

\subsection{Chandra}
The X-ray observations in table~\ref{tab:obslog} were performed with
\emph{Chandra}/ACIS-S in the VFAINT mode. HD~163296 is placed on the back-illuminated
chip S3, because this chip has the best sensitivity for soft X-rays. This ACIS chip has a spatial resolution of 0.492\arcsec{} per pixel and an intrinsic energy resolution of 110-140~eV for energies below 2~keV. 
We reprocessed all \emph{Chandra} data with CIAO~4.4 \citep{2006SPIE.6270E..60F}. Light curves and spectra for the central source are extracted using standard CIAO scripts. The background was determined from a large, source-free region on the same chip. Spectral fitting was done with the SHERPA fitting tool \citep{2007ASPC..376..543D}.

The ACIS point-spread function (PSF) is under-sampled by the pixel size. To avoid an aliasing on the pixel grid two techniques can be used. Either the photon positions are randomized with offsets up to half a pixel or their positions are corrected with an energy-dependent sub-pixel event repositioning \citep[EDSER,][]{2004ApJ...610.1204L}. The first method leads to a well-characterized, but artificially broadened PSF; the second method is less well characterized and was implemented in CIAO only recently, but it increases the resolution of the resulting images directly. All images shown here are processed with the EDSER algorithm, but we verified that images with randomized pixel positions give compatible results.

The background light curves for both observations are flat confirming that no flaring of the ambient spacecraft proton impact rate was present.

\subsection{HST/STIS}
The jet of HD~163296, HH~409, has been observed with \emph{HST} in different instrument configurations. Here, we analyze archival and new STIS long-slit spectroscopy, where the slit is aligned with the jet axis. The observations cover two different wavelength ranges: The \texttt{G140M} grating is used for FUV spectroscopy, most importantly the hydrogen Ly$\alpha$ line, and the \texttt{G750L} and \texttt{G750M} gratings provide information on optical FELs with different wavelength resolutions. Table~\ref{tab:obslog} gives details of the datasets used. In most cases the slit is narrow (52\arcsec{}$\times$0\farcs2), thus the exact position angle is important, only observation O66Q02010 was taken with a wider slit (0\farcs5 versus 0\farcs2). \citet{2000ApJ...544..895G} determine a jet orientation of $42.5\pm3.5$\degr. 

HD~163296 is optically bright ($m_V = 6.9$~mag). To avoid saturation of the CCD the central star is placed behind the coronagraphic wedge (width 0.5\arcsec{}, aperture \texttt{52X0.2F1}) in all optical spectra. The same setup was chosen for our new FUV spectrum as well. 

\subsubsection{New optical observations}

\onlfig{
\begin{figure*}
\includegraphics[width=10cm]{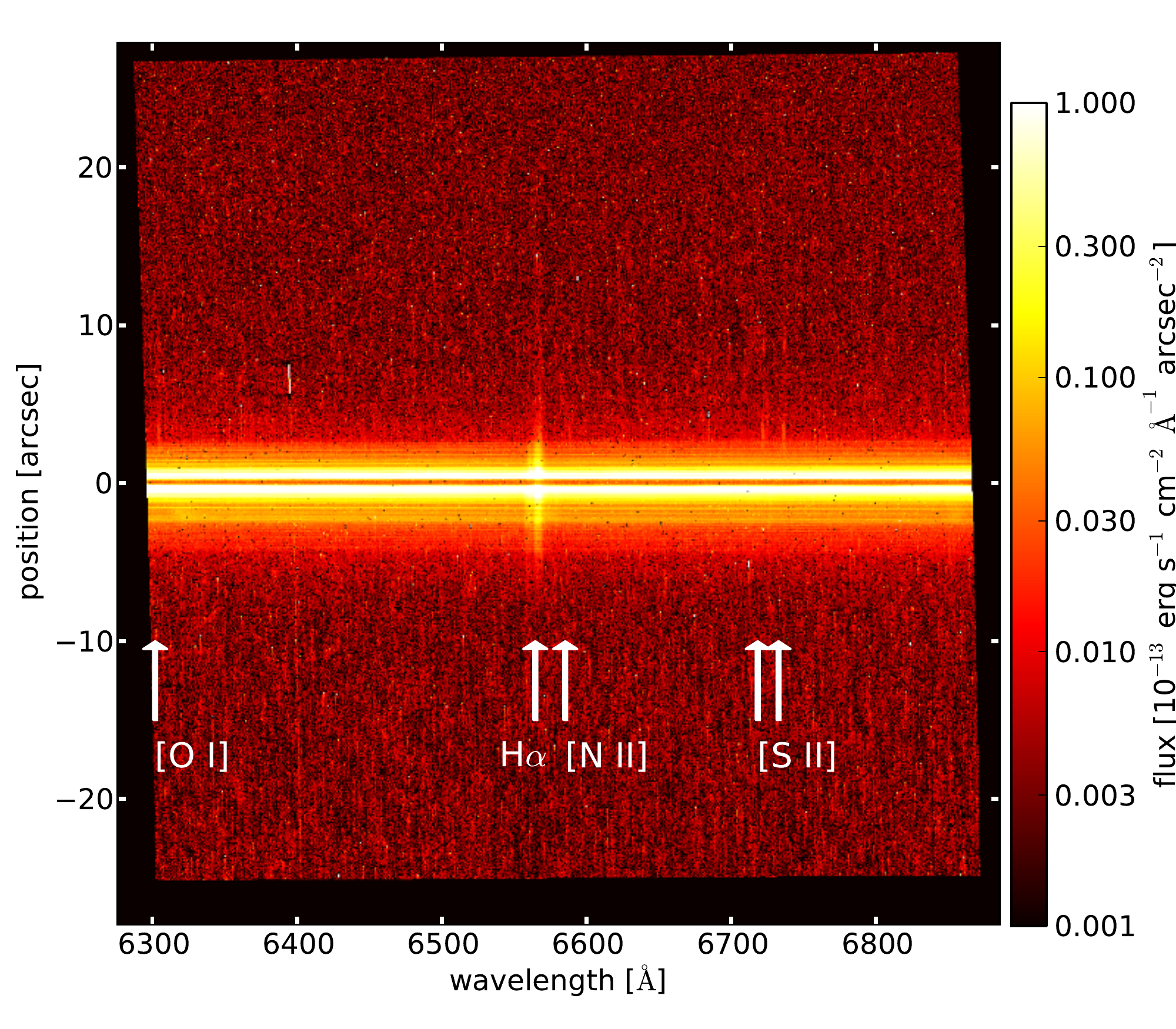}
\caption{G750M data from 2011. Hot pixels close lines of interest have been removed manually (see text for details). Lines which we find to be spatially extended (see Sect.~\ref{sect:opticalemission}) are marked. \label{fig:fullimageOpt}}
\end{figure*}
}
Pipeline processed STIS data were downloaded from MAST. A full detector image is shown in Fig.~\ref{fig:fullimageOpt}.
Optical CCD images are corrected for overscan and bias, then dark-subtracted, flat-fielded, rectified and wavelength calibrated to a heliocentric reference frame. The spectra span the wavelength range from approximately 6300-6865~\AA{} and the resolving power is $R\approx 5000$.

Despite the coronagraphic wedge, there is a significant stellar contribution in the optical data (Airy rings). Below we show position-velocity diagrams (PVDs) of the strongest lines in our data. To remove the stellar contribution we define two wavelength intervals to the left and to the right of the emission line of interest (The specific wavelength regions chosen are given with the PVD). We then estimate the stellar contribution in the region of the PVD by a linear interpolation between the left and the right wavelength region. This procedure works because the wavelength intervals of interest are very narrow. However, low surface brightness features in the jet cannot be identified close to the central object due to the low contrast to the bright Airy rings.
Furthermore, the strong, presumably stellar, H$\alpha$ profile scatters into the outer jet regions and appears as extra emission in the continuum-subtracted position-velocity diagrams. A comparison with spatial profiles close to the H$\alpha$ wavelength reveals that all excess flux is compatible with scattering from the central position.

\subsubsection{New FUV observations}
Due to a change in the Science Data Formatter (SDF) on HST not all events in the new G140M exposure were transferred from the SDF to the on-board solid-state recorder (SSR) for downlink. This leads to a gap of 138.5~s in the exposure. Once the damaged period is removed from the telemetry stream, the resulting raw data is processed with the standard calstis pipeline in the same way as the raw data from all other observations. A full detector image is shown in Fig.~\ref{fig:fullimageFUV2011}.
\onlfig{
\begin{figure*}
\includegraphics[width=10cm]{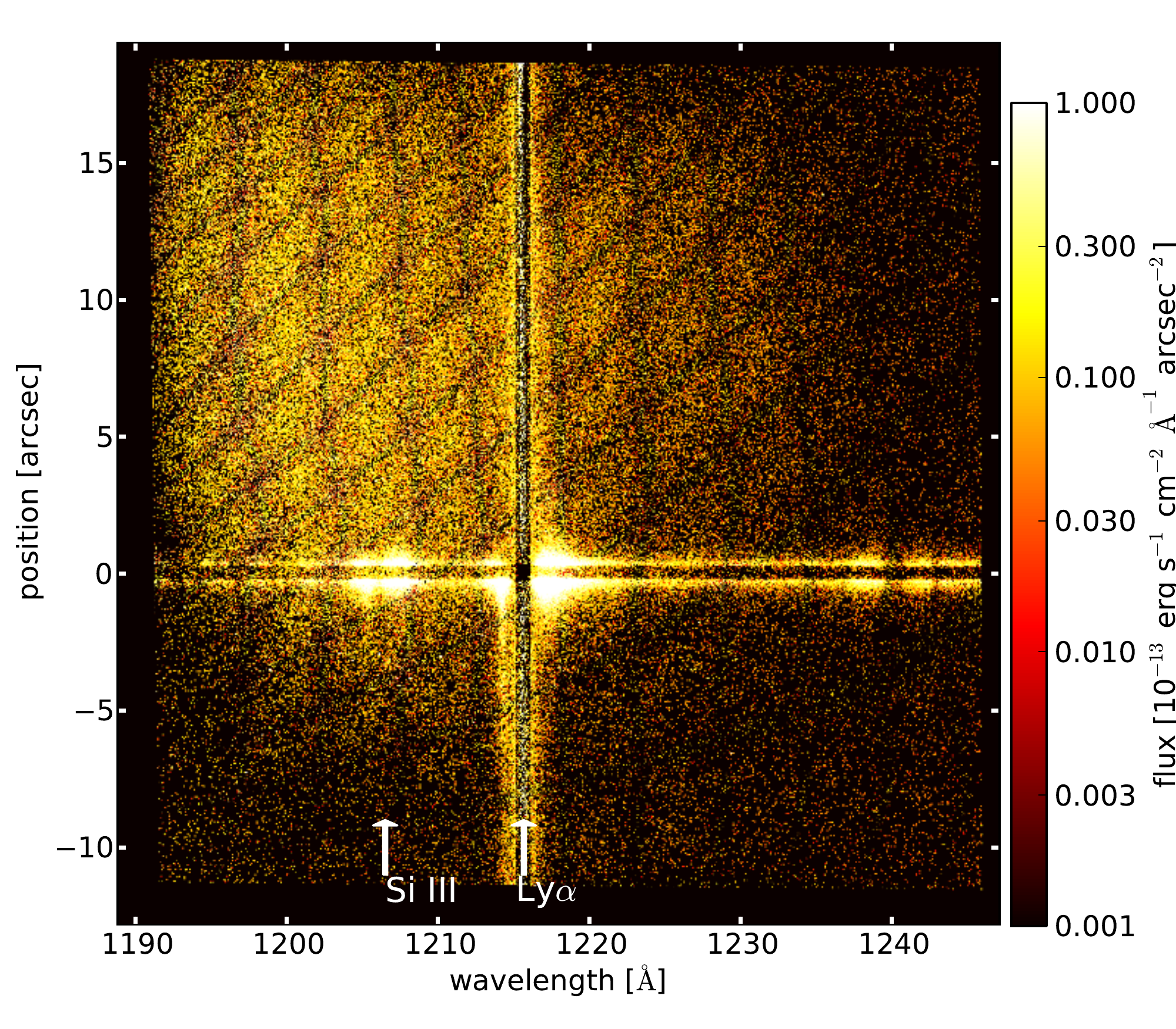}
\caption{G140M data from 2011. The Ly$\alpha$ line airglow feature has been removed. Lines which where found to be extended in earlier observations  are marked.\label{fig:fullimageFUV2011}}
\end{figure*}
}

CalSTIS divides the image by a flat-field, then it is rectified and wavelength calibrated. The data covers approximately the range 1190-1245~\AA{}. We use a binning of 4 high-resolution pixels, which provides a wavelength resolution of 0.2~\AA{} for a point source. For comparison the effective resolution of emission that fills the slit uniformly, would be $\approx80$~km~s$^{-1}$.

While the overall dark current is low, the top left hand corner of the FUV-MAMA detector shows an enhanced glow that is intrinsic to the micro-channel plate array. The strength of the glow is time variable, but it can be removed by scaling archival dark exposures. However, these have a limited signal-to-noise ratio and thus the subtraction would increase the noise in the residual image. Since the glow changes on much larger spatial scales than the features of the jet, we fit the local background around the knots of the jet instead of subtracting it globally.

The FUV images are dominated by strong Ly$\alpha$ airglow emission. For each position we determine the centroid of the airglow line. We build an average line profile from regions close to the detector edges and far from the known jet emission. At each position we subtract this profile from the image. The accuracy of this procedure is limited by the statistical noise in the airglow line. We estimate that only features with a surface luminosity of $>10$\% of the observed airglow line can be detected. All residuals are smaller than this value. The data is taken in \texttt{TIME-TAG} mode, and as a cross-check we limited the analysis to times where the count rate in the airglow is less than a third of the maximum rate. We conclude that no astrophysical signal is found at the position of the airglow line.

In the FUV the Airy rings are much less severe. The generally lower count rates mean that the interpolation is not as well defined and the procedure described above would introduce significant noise to the PVDs. Therefore, no subtraction of the contribution from the unresolved source was attempted for the FUV exposures.

\subsubsection{Archival data}
Pipeline processed STIS data were downloaded from MAST and treated in the same way as the new FUV or optical data.
While all G140M observations are useful to analyze material close to the central source, we use only the observations from 1999 and 2011
when we look at the knots of the jet, which are located a few arcsec from the star
because \citet{2006ApJ...650..985W} show that the position angle of 46.5\degr{} in the deeper observation O66Q03010 is too large and the slit misses the apex of the jet shocks. Observation O66Q02010 is the only dataset taken with a wider slit. While knots are visible, their flux is uncertain because they partially overlap with the strong Ly$\alpha$ airglow.

\subsubsection{Data reduction}
We further processed the data with custom python routines. Hot or negative pixels are marked by hand and replaced with the average value of the surrounding pixels. The jet emission is weak both in the FUV and in the optical. Consequently, the signal-to-noise ratio (SNR) for a single pixel is typically 2-4 in the jet. However, all features analyzed stretch over a larger detector area. We present the data binned in the spatial direction over 4 pixels for the FUV and over 6 pixels for [\ion{O}{i}] 6300\AA{} and the [\ion{S}{ii}] lines at 6716\AA{} and 6731\AA{}. When we fit fluxes and positions for knots in the jet, we achieve even better SNR because they stretch over a range of velocities as well. Uncertainties given in the tables are $1\sigma$ confidence intervals throughout the article.

\section{Results}
\label{sect:results}

\subsection{X-ray emission}
The first \emph{Chandra} observation of HD~163296 was already presented by \citet{2005ApJ...628..811S} but we analyze it here again for comparison. 
HD~163296 was also observed with \emph{XMM-Newton} by \citet{HD163296}.

\subsubsection{The central star}

The light curves and hardness ratios for the archival observations are
moderately variable and the new observation is fully compatible with
those findings. The \emph{XMM-Newton} spectra have a higher
signal-to-noise ratio than the \emph{Chandra} data because the
effective area of \emph{XMM-Newton} is larger, the exposure time is
longer and \emph{XMM-Newton}/RGS delivers a high-resolution grating
spectrum. We use the same spectral model for HD~163296 as in \citep{HD163296}
(one \texttt{phabs} absorber and three \texttt{vapec}, optically thin,
collisionally excited, thermal emission components with non-solar
abundances) and keep the abundances fixed at the values found in the
\emph{XMM-Newton} observation. Elements like Mg, Fe, Si and C with a
low first-ionization potential (FIP) are enhanced compared to O and Ne
which have a higher FIP. This is a common effect in inactive
stars. Also, we keep the absorbing column density fixed at $N_{\rm H}
= 7 \times 10^{20}$~cm$^{-2}$, which is in full agreement with the
optical reddening \citep{HD163296}. 

Table~\ref{tab:3vapecem} lists the temperature and volume emission
measures for the three temperature components. Both \emph{Chandra}
observations are refit here, the reduced $\chi^2$ values of both fits
are below 1. Statistically acceptable fits are possible with two temperature components only, but we keep the model of \citet{HD163296} to facilitate a comparison. The fit values for the \emph{XMM-Newton} observation are taken directly from that work (after correction for the revised distance we use in this article).
\begin{table}
\caption{\label{tab:3vapecem}Best-fit model parameters (1$\sigma$ confidence interval).}
\centering
\begin{tabular}{lllll}
\hline \hline
\multicolumn{2}{c}{component} &  soft & medium & hard \\
\hline
\multicolumn{5}{c}{Parameters fixed for all components\tablefootmark{a}}\\
\hline
$N_H$ & [$10^{20}$~cm$^{-2}$] & 7 & 7 & 7\\
C & \tablefootmark{b}& 3.7 & 3.7 & 3.7\\
O & \tablefootmark{b}& 0.7 & 0.7 & 0.7\\
Ne& \tablefootmark{b}& 1.2 & 1.2 & 1.2\\
Mg& \tablefootmark{b}& 2.3 & 2.3 & 2.3\\
Si& \tablefootmark{b}& 2.8 & 2.8 & 2.8\\
Fe& \tablefootmark{b}& 1.6 & 1.6 & 1.6\\
\hline
\multicolumn{5}{c}{\emph{Chandra} 2003 (red. $\chi^2=0.91$)}\\
\hline
k$T$ & [keV] & $0.19^{+0.23}_{-0.02}$ & $0.60^{+0.05}_{-0.04}$ & $2.0^{+1.0}_{-0.5}$ \\
$VEM$ & [$10^{52}$~cm$^{-3}$] & $1.3^{+0.5}_{-0.5}$ &
$0.9^{+0.2}_{-0.2}$ & $0.4^{+0.1}_{-0.1}$\\
\hline
\multicolumn{5}{c}{\emph{XMM-Newton} 2007\tablefootmark{c}}\\
\hline
k$T$ & [keV] & $0.21^{+0.02}_{-0.01}$ & $0.51^{+0.1}_{-0.02}$ & $2.7^{+1.0}_{-0.5}$ \\
$VEM$ & [$10^{52}$~cm$^{-3}$] & $2.2^{+0.5}_{-0.3}$ & $1.1^{+0.3}_{-0.4}$ & $0.5^{+0.1}_{-0.2}$\\
\hline
\multicolumn{5}{c}{\emph{Chandra} 2011 (red. $\chi^2 = 0.85$)}\\
\hline
k$T$ & [keV] & $0.16^{+0.01}_{-0.01}$ & $0.60^{+0.2}_{-0.02}$ & $2.4^{+0.5}_{-0.3}$ \\
$VEM$ & [$10^{52}$~cm$^{-3}$] & $1.3^{+0.2}_{-0.2}$ &
$1.0^{+0.6}_{-0.6}$ & $0.4^{+0.1}_{-0.1}$\\
\hline
\end{tabular}
\tablefoottext{a}{from the \emph{XMM-Newton} fit. See
  \protect{\citet{HD163296}} for details.}
\tablefoottext{b}{abundances relative to the solar values of
  \protect{\citet{1998SSRv...85..161G}}}
\tablefoottext{c}{All values taken from \protect{\citet{HD163296}}.}
\end{table}
The temperatures of all three components are very similar and consistent within the errors over all three observations. The same is true for the volume emission measure ($VEM$) with the exception of the soft component, which seems marginally stronger in the \emph{XMM-Newton} observation.

The light curves show little variability during the observations and
the comparison of the spectra and their emission measures over nearly
one decade shows that the X-ray emission of the HAeBe HD~163296 is
stable. The observed (absorbed) energy flux in the 2011 observations is
  $(2.4\pm0.2)\times10^{-13}$~erg~s$^{-1}$~cm$^{-2}$ (0.3-5.0~keV).
From this we derive an intrinsic $\log L_X = 29.8$ (in
  ergs~s$^{-1}$), which is almost identical to the values of previous
  observations \citep{HD163296}.

\subsubsection{The base of the jet}
\label{sect:basejet}
Based on a line ratio in the \ion{O}{vii} He-like triplet that is sensitive to the ambient UV field \citet{HD163296} conclude that the soft X-ray emission must be emitted at least $0.8\;R_*$ above the stellar surface.

In the CTTS DG~Tau \citet{Schneider} directly measure an offset of about 40~AU between the centroids of the soft and the hard component of the central source. The hard component is presumably stellar emission, while the soft component might originate in a jet shock close to the star. This detection is possible in DG~Tau, because the high column density absorbs all soft X-rays from the star itself. We search for an offset using the same technique in HD~163296 but find no hint of extension. A larger physical separation would be required in HD~163296 than in DG~Tau, because the absorbing column density is very low and so any stellar plasma would also contribute to the soft band flux.

Additionally, we simulate the \emph{Chandra} point-spread function with SAOTrace Version 2.0.1 \citep{2003ASPC..295..477C}. This tool performs Monte-Carlo ray-tracing through the \emph{Chandra} optics. In the current version it includes the telescope dither pattern. From the ray-trace we calculate the detector response with MARX \citep{1997ASPC..125..477W}, version 5.0. The combination of SAOTrace and MARX can properly simulate the effect of the EDSER sub-pixel event repositioning. We find an extension beyond the simulated PSF in the range 0.5-0.8\arcsec{}. However, a deconvolution of the observed image with the simulated PSF using the CIAO tool \texttt{arestore} shows that the extension in both observations is perpendicular to the jet direction. It matches the PSF asymmetry recently detected in \emph{Chandra}/HRC observations \citep{2010HEAD...11.4011J} in both the distance to the center of the PSF and in the relative position in detector coordinates. Thus, the central source itself is apparently not extended. In comparison to DG~Tau, where the jet is detected within 30~AU of the central star using image deconvolution (G\"udel et al., in prep.) we conclude that almost all soft X-ray emission from HD~163296 is contained within the inner 30~AU.

\subsubsection{The X-ray knots of the jet}

\begin{figure*}
\sidecaption
\includegraphics[width=6cm]{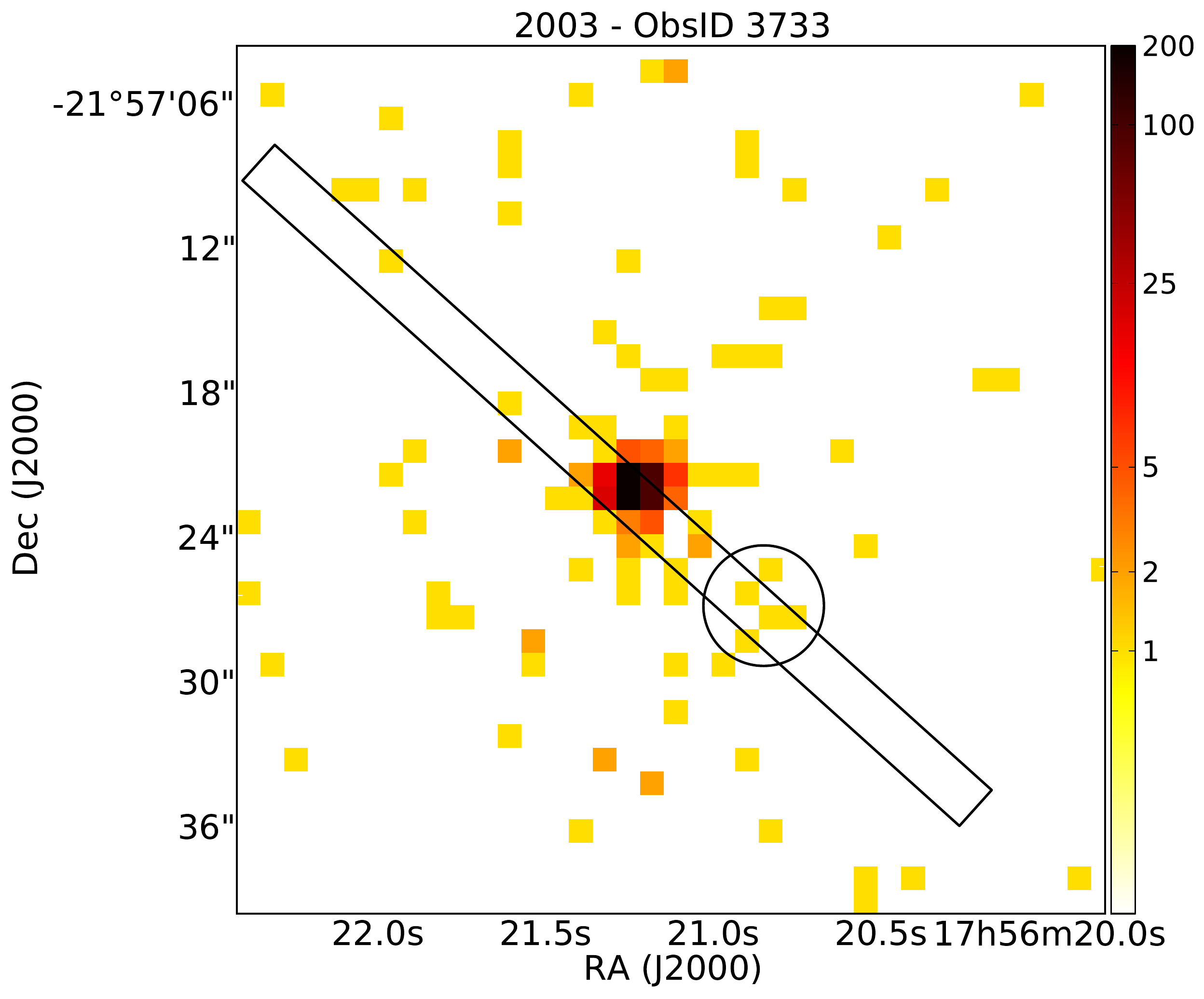}
\includegraphics[width=6cm]{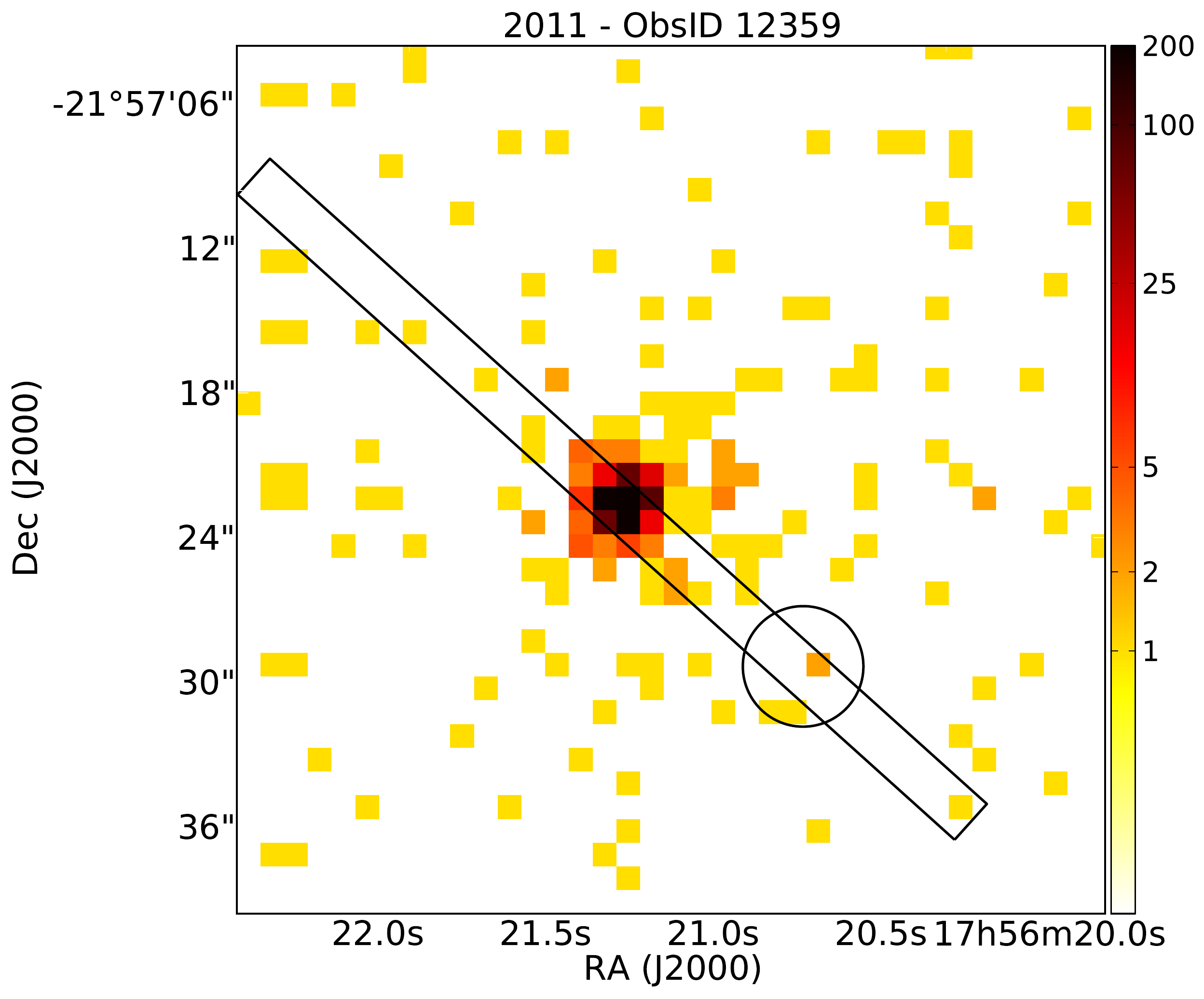}
\caption{\emph{Chandra}/ACIS image of HD~163296 and its jet HH~409 in
  the energy range 0.2-5.0~keV binned up by a factor of two (Each
  block is $2 \textnormal{ pixel}\approx1$\arcsec{} on the side.). The black box  shows the position of
  the optical jet. The half-width of the box is comparable to the PSF
  (1\arcsec). The long axis of the box passes through the
  centroid of the X-ray emission and has the position angle 42\degr. Circles indicate locations of possible emission knots. The radii are 2.5\arcsec{}. See the electronic edition of the Journal for a color version 
of this figure. \label{fig:counts_image}}
\end{figure*}

\begin{figure}
\resizebox{\hsize}{!}{\includegraphics{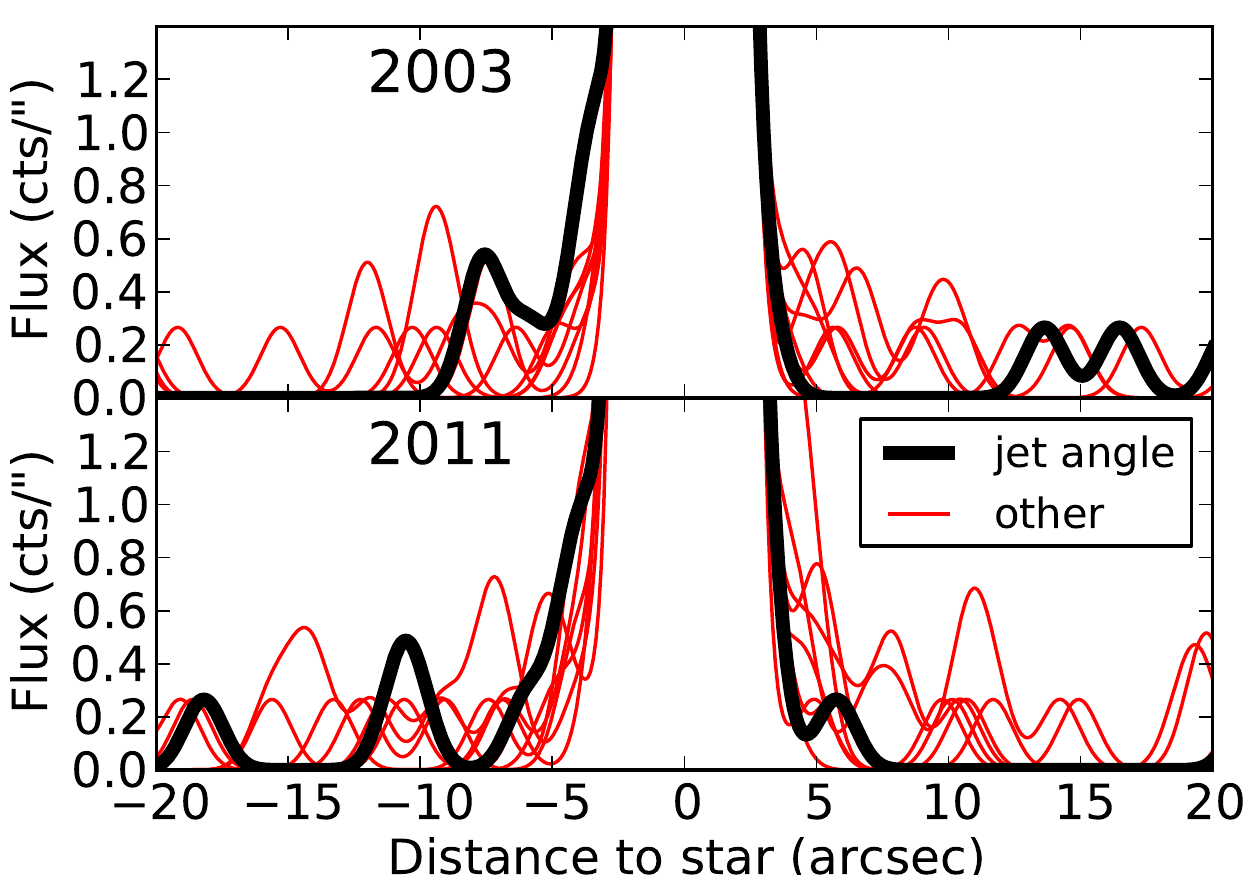}}
\caption{Black line: Projection of \emph{Chandra}/ACIS counts onto the jet axes in a rectangle with half-width 1\arcsec{}. red/gray lines: projection onto a set of lines through HD~163296, but with different angles. All fluxes are smoothed. Distances are positive for the NE jet and negative for the SW jet. See the electronic edition of the Journal for a color version 
of this figure. \label{fig:jet_projection}}
\end{figure}

Figure~\ref{fig:counts_image} shows the region around HD~163296 in X-rays. 
The position angle of the jet is well known from optical images. Thus, we can select the jet region \emph{a-priori} and do not need to rely on a source detection in the \emph{Chandra} images. We select events with energies in the range 0.2-2.0~keV. Fig.~\ref{fig:jet_projection} shows a projection of all counts within 1\arcsec{} of the jet axis. 
The black lines show the count distribution along the jet. For comparison, the same method was employed for a set of angles rotated by 20\degr{}, 40\degr{}, 60\degr{}, 80\degr{}, 100\degr{}, 120\degr{}, 140\degr{} and 160\degr{} with respect to the jet (red lines).

In the data from 2003 the black line is above the comparison lines in the range -3 to -4\arcsec{} (negative numbers are for the approaching south-west jet). To estimate the significance, we determine an average number of 1.75 counts between -4.5\arcsec{} and -2.5\arcsec{} from the wings of the PSF, while 5 counts are observed in this range along the jet direction. In a Poisson distribution the chance to observe 5 counts when the expectation value is 1.75 is 3\%. Thus, this feature is 97\% significant. Note, that this does not contradict the results of Sect.~\ref{sect:basejet}, because this feature is two orders of magnitude less luminous than the central star and thus it does not show up in an image deconvolution. This emission corresponds to the extension of the source seen by \citet{2005ApJ...628..811S}.
A similar feature in 2011 is seen in the black line but also in directions other than the jet (red lines in the figure), thus it is not clear if this feature is present in 2011.

In the projection the strongest features beyond the stellar PSF are two peaks at -7.2\arcsec{} and -10.6\arcsec{} in 2003 and 2011, respectively. 
\citet{2005ApJ...628..811S} derive a probability of only 0.04 that the
knot in 2003 (black circle in the left panel of Fig.~\ref{fig:counts_image}) 
is a chance fluctuation by calculating the number of circles of this
size at any position on the detector that contain 5 or more
photons. We confirm this number using Poisson statistics of the
measured background in a near-by source free region (0.008 and
0.012~cts~pixel$^{-1}$ in 2003 and 2011, respectively, in the energy
range 0.2-2.0~keV). Thus, we expect 0.65 and 0.97 background events in
a circle with radius 2.5\arcsec{}. According to Poisson statistics the
chance to find 5 or more photons in any one circle in 2003 is
therefore $<0.1$\% and the chance is 7\% to find 3 or more photons in 2011. To estimate the significance of the detection, we need to consider the number of circles searched: We analyze a region up to 25\arcsec{} from the central star in both jet directions, which can be covered by ten circles with radius 2.5\arcsec{}. (We would recognize features that fall between two of those circles, thus more than ten positions are searched. However, only the number of \emph{independent} positions is important.) 
In 2003, the probability to find no circle with 5 counts or more is 99\%, which we can take as the significance of the detection of an X-ray feature in the jet; the significance of an independent detection of the X-ray feature in the jet in 2011 is only 44\%.  
A potential knot with three photons in 2011 is not significant by itself.
However, if we assume the X-ray feature in 2003 is real and then propagate it with the proper motion of the optical knot~A, which shows $0\farcs49\pm0\farcs02$~yr$^{-1}$ \citep{2006ApJ...650..985W}, then this defines the search area exactly (circle in Fig.~\ref{fig:counts_image}, right panel) 
and the chance to find three of more photons in a single circle of this size is $<7$\%. Thus, we can say that \emph{assuming the knot in 2003 is real and has the same proper motion as the optical knots}, we detect an X-ray knot in 2011 with 93\% significance.

We note, that the half-width of the detection region used by \citet{2005ApJ...628..811S} is 2.5\arcsec{}, much larger than expected from the width of the optical knots and the \emph{Chandra} PSF. Fig~\ref{fig:jet_projection} shows the projection of the counts on the jet axis with an extraction region matched to the PSF. Several features in the comparison sample (projected on lines which do not coincide with the jet axis) have similar fluxes as the tentative knots along the jet. However, in HH~2 \citet{2012A&A...542A.123S} find that the X-ray emitting region is larger than the optical knots. If the same is true here, then it is not surprising, that the knots are only detected if we use a large extraction region.

The effective area of the \emph{Chandra}/ACIS detector degraded
between 2003 and 2011 (about 5\% at 1~keV, 20\% at 0.8~keV and 40\% at
0.5~keV), but the observation taken in 2011 is more than twice as
long. Taking the three photons at -10.6\arcsec{} as an upper limit to
the flux of the tentative X-ray knot, the apparent luminosity
decreased by a factor three compared to 2003. Although the count
number is low the hypothesis that the photon flux is the same in both
observations can be rejected on the 95\% level. 

In summary, we can say that in 2003 an X-ray knot is detected with 95\% significance, confirming \citet{2005ApJ...628..811S}. Under the assumption that this knot moves with a proper motion similar to the optical knots \citep[as observed in HH~2][]{2012A&A...542A.123S}, this feature is detected again in 2011, but with lower luminosity.

\subsubsection{Summary of X-ray results}
HD~163296 itself is very stable over the last decade. Lightcurves and
hardness ratios in all observations are nearly flat and the
temperatures and emission measures of all spectral fits agree. From
the \ion{O}{vii} line ratio \citet{HD163296} conclude, that the soft
X-ray emission must originate above the stellar surface, but we do not
see an offset between the soft and the hard centroid, setting an upper
limit on the separation of 30~AU. However, very weak emission features
(knots of the jet) are seen on the edge of the PSF at -4\arcsec{}
along the jet. Furthermore, there is a  weak feature at -7.2\arcsec{}
in 2003 and a tentative feature at -10.6\arcsec{} in 2011
(table~\ref{tab:knotpos}). Both of these features do not have
  an optical or UV counterpart (see below).

\begin{table}
\caption{Position of knots from Gaussian fit\label{tab:knotpos}}
\centering
\begin{tabular}{lllr}
\hline\hline
knot & tracer & epoch & position\\
\hline
X-ray knot & X-ray & 2003.69 & $-7.2\pm1.0$\arcsec{}\tablefootmark{a}\\
X-ray knot & X-ray & 2011.11 & $-10.6\pm0.3$\arcsec{}\tablefootmark{a}\\
       & Ly$\alpha$ & 2011.64 & $\approx -2$\arcsec{}\\
       & Ly$\alpha$ & 2011.64 & $-4.0\pm0.2$\arcsec{}\\
knot B & [\ion{S}{ii}] 6731\AA{} & 1999.69 & $4.3\pm0.1$\arcsec{}\\
knot B  & [\ion{S}{ii}] 6731\AA{} & 2011.64 & $8.7\pm0.1$\arcsec{} \\
knot B2  & [\ion{S}{ii}] 6731\AA{} & 2011.64 & $3.3\pm0.1$\arcsec{} \\
\hline
\end{tabular}
\tablefoottext{a}{The uncertainty is calculated from the variance of
  the photon position for all photons attributed to the knot.}
\end{table}

\subsection{UV emission}
HH~409 was observed at UV wavelengths with STIS previously in 1999
and in 2000. 
The most prominent line in the FUV is Ly$\alpha$. Its central part is contaminated by ISM absorption and airglow, which fills the aperture. Thus, only data more than 150~km~s$^{-1}$ from the line center can be used. \citet{2000ApJ...542L.115D} found blue-shifted Ly$\alpha$ emission in the jet with a velocity around $380$~km~s$^{-1}$ at 0.06\arcsec{} (7.1 AU), that decreased to $335$~km~s$^{-1}$ at 6\arcsec{} (714 AU). Additionally, they could identify three knots (Knot A in the jet at -7.7\arcsec{} and knot B and C in the counter-jet at 4.5\arcsec{} and 8.6\arcsec{}, respectively).

\begin{figure*}
\sidecaption
\includegraphics[width=12cm]{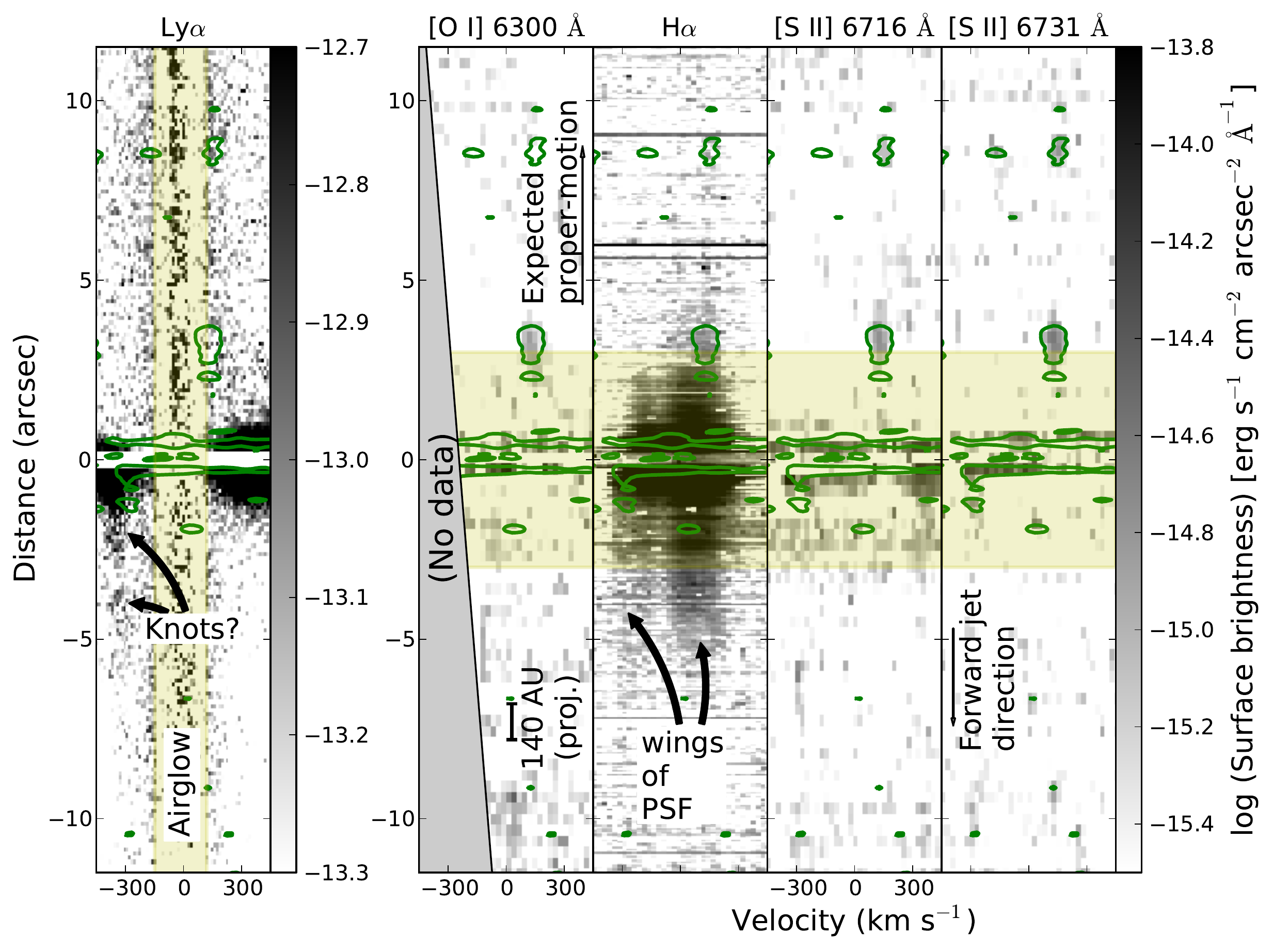}
\caption{PVD for Ly$\alpha$, H$\alpha$ and three optical FELs. For comparison, the contourlines for [\ion{S}{ii}]~6731\AA{} are overlaid in all panels.
\emph{left:} The Ly$\alpha$ line is contaminated by airglow close to the rest velocity. \emph{right:} The optical lines show scattered light close to the central star despite the occulting bar. These areas are shaded. [\ion{O}{i}] 6300\AA{} is located close to the edge of the chip, so no data exists in the bottom left of the PVD. Contour lines are drawn for [\ion{S}{ii}]~6731\AA{} at the level $7\times10^{-16} $erg~s$^{-1}$~\AA{}$^{-1}$~cm$^{-2}$~arcsec$^{-2}$. The optical PVDs are background subtracted (see text). The following wavelength intervals are used to fit the background: [\ion{O}{i}] 6300\AA{}: 6352-6357\AA{} and 6376-6382\AA{}; H$\alpha$: 6504-6524\AA{} and 6604-6624\AA{}; [\ion{S}{ii}]~6716\AA{}: 6705-6708\AA{} and 6721-6724\AA{}; [\ion{S}{ii}]~6731\AA{}: 6721-6723\AA{} and 6735-6738\AA{}.
See the electronic edition of the Journal for a color version 
of this figure. \label{fig:PVD}}
\end{figure*}

\begin{figure*}
\includegraphics[width=0.49\textwidth]{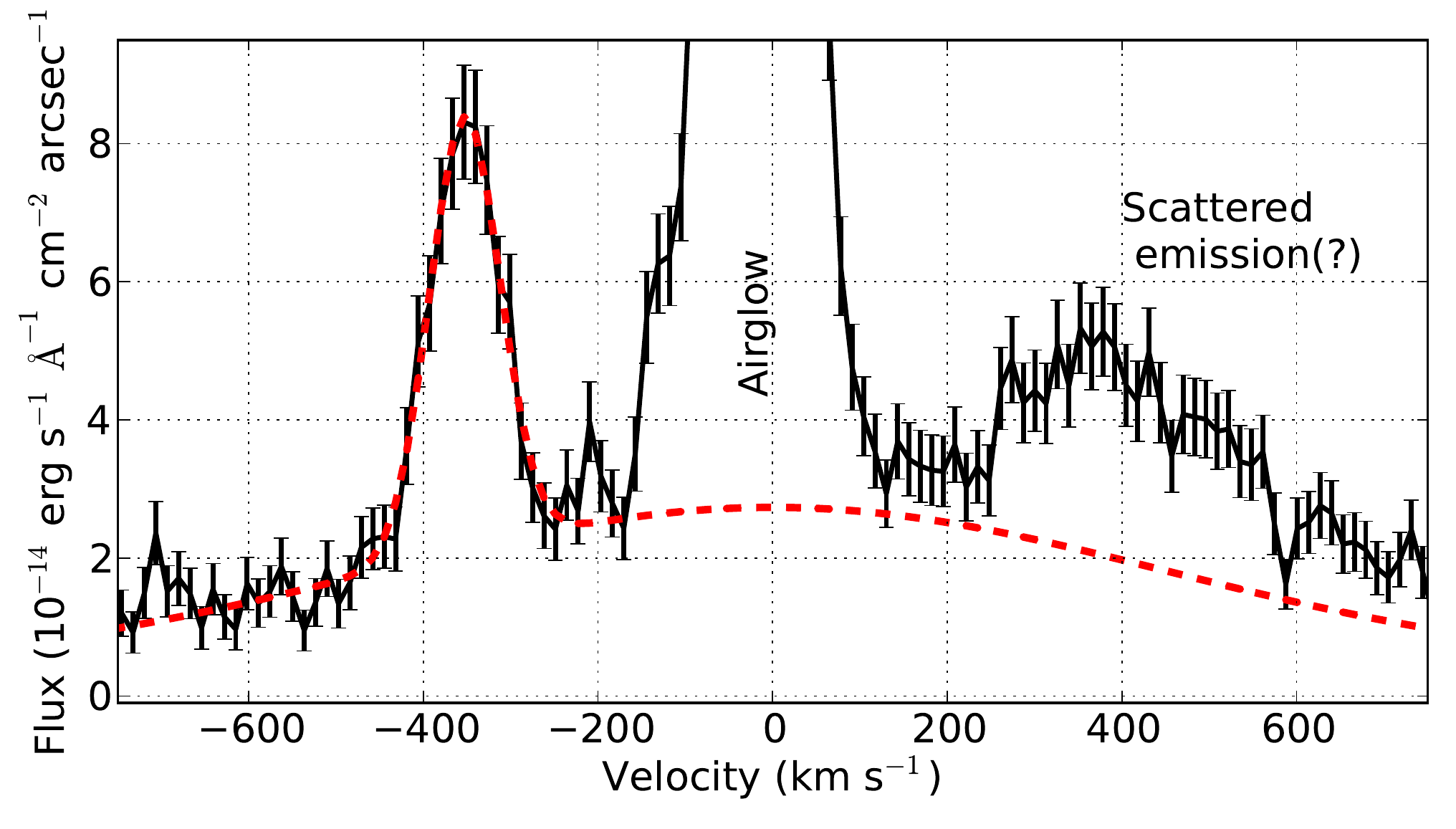}
\includegraphics[height=0.49\textwidth,angle=90]{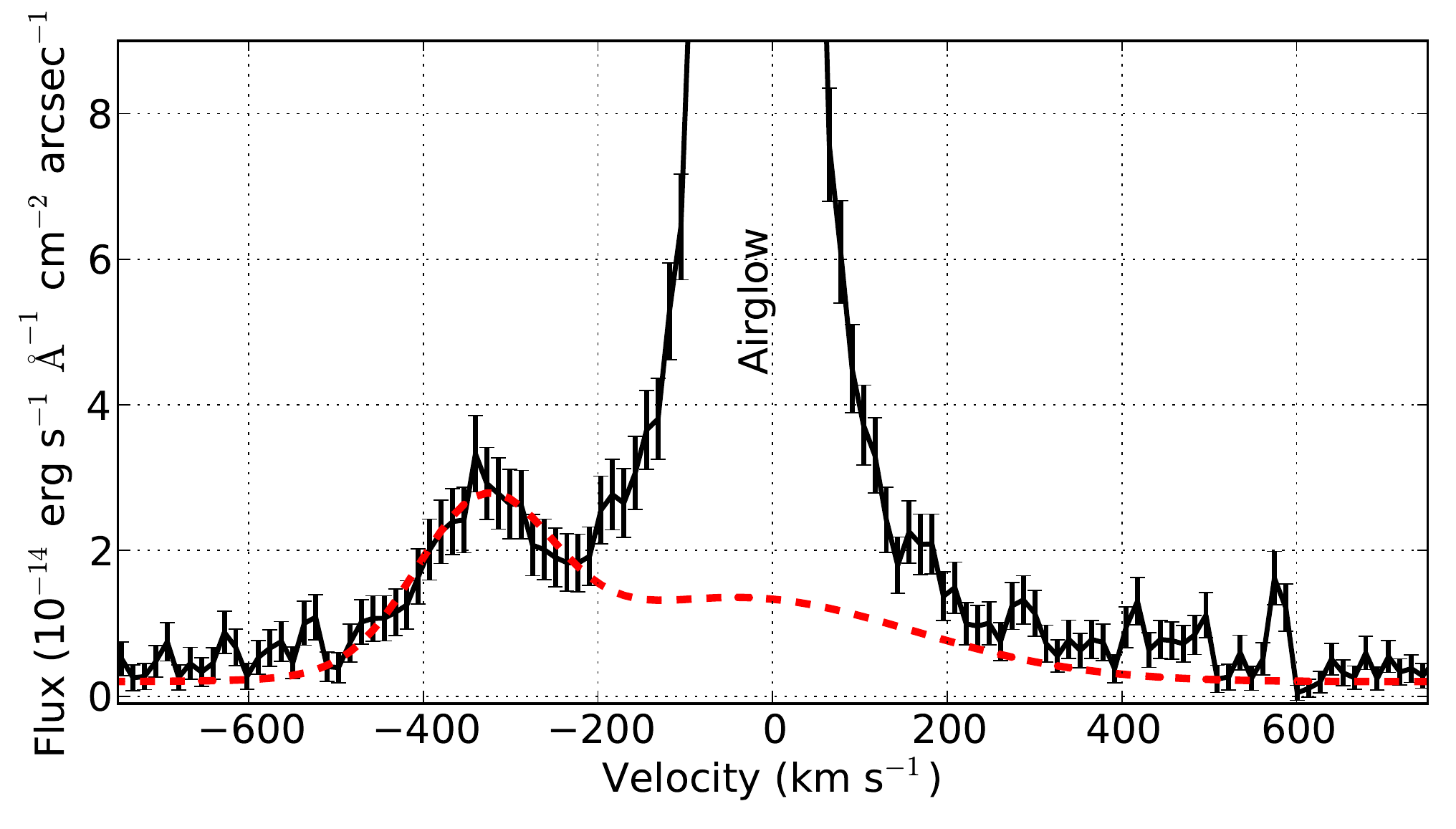}
\caption{Spectra of the Ly$\alpha$ line in the two brightest knots. The inner part of the line is contaminated by airglow, additionally the knot closer to the source (left panel) shows some emission on the red side of the line which could be due to scattered light from the central star (see text). We fit two Gaussians for knot and background (dashed line). Both knots appear blue-shifted by about 350~km~s$^{-1}$. \label{fig:FUVspec}}
\end{figure*}

Figure~\ref{fig:PVD} (left) shows the position-velocity diagram (PVD) of Ly$\alpha$ for the new observation. Despite the slightly longer exposure time the data quality of the new observation does not reach the level of the observation in 1999. 

In the jet itself there are two knot features at about -2\arcsec{}  and -4\arcsec{}. At the given SNR ratio they are well described by a Gaussian. Fitted values for the position are given in table~\ref{tab:knotpos}, line fluxes in table~\ref{tab:fluxes}. Figure~\ref{fig:FUVspec} shows spectra extracted at the position of the two knots. This is the same range of velocities and speeds observed by \citet{2000ApJ...542L.115D} and indicates that the flow speed of the jet has not changed considerably in the last decade.
 
The \ion{Si}{iii} line at 1206.5~\AA{} is too weak to be fitted independently. Thus, we fix the line shift and width at the values found for Ly$\alpha$. For the outer knot we can only give an upper limit, for the inner one we formally find a detection (table~\ref{tab:fluxes}). However, we caution that the flux is weak and the position of the inner knot coincides with a region of enhanced detector background in \ion{Si}{iii}.
Thus, the flux might be overestimated.

In principle the models of \citet{1987ApJ...316..323H} constrain the shock velocity given the Ly$\alpha$/\ion{Si}{iii} flux ratio. These model are based on simulations of radiative shocks. The pre-shock material is assumed to be in ionization equilibrium with the post-shock material. This describes a scenario where the Lyman radiation from the shock front is the main ionization source for the pre-shock gas. At a distance of a few hundred AU from the central star this is a reasonable assumption.

In all these models the flux ratio Ly$\alpha$/\ion{Si}{iii}$> 10$. Thus,
they are not constrained by the upper limit in the outer knot and they cannot explain the tentative \ion{Si}{iii} detection in the inner knot. It is possible that scatter from the central star contributes to the inner emission similarly to the spatially extended Ly$\alpha$ emission, which is seen red-shifted on the side of the approaching jet and which possibly also contributes to the blue-shifted emission.
 
The observed surface brightness of the Ly$\alpha$ emission is reduced
compared to 1999. This can be seen in table~\ref{tab:fluxes}
  that shows the fluxes integrated over the visible knot surface.
Either the jet is dimmer in 2011 or it has broadened considerably, so that only part of the total luminosity is contained in the aperture. The later explanation seems unlikely, because the jet was centrally bright in previous observations. Also, in a quasi-static scenario where gas travels along the jet but the overall shape of the jet stays constant with time this is not expected.

The line fluxes in table~\ref{tab:knotpos} are not dereddened. \citet{2000ApJ...542L.115D} found $E(B - V) = 0.015$, which leads to flux correction of 15\% \citep{2006ApJ...650..985W}. 

Given the proper motion of the old knots A and C \citep{2006ApJ...650..985W} they should be located at -13.6\arcsec{} and 12.7\arcsec{} during the new observation. No significant emission is observed at these locations, nor at the tentative X-ray knot at -10.6\arcsec{}.

\begin{figure}
\resizebox{\hsize}{!}{\includegraphics{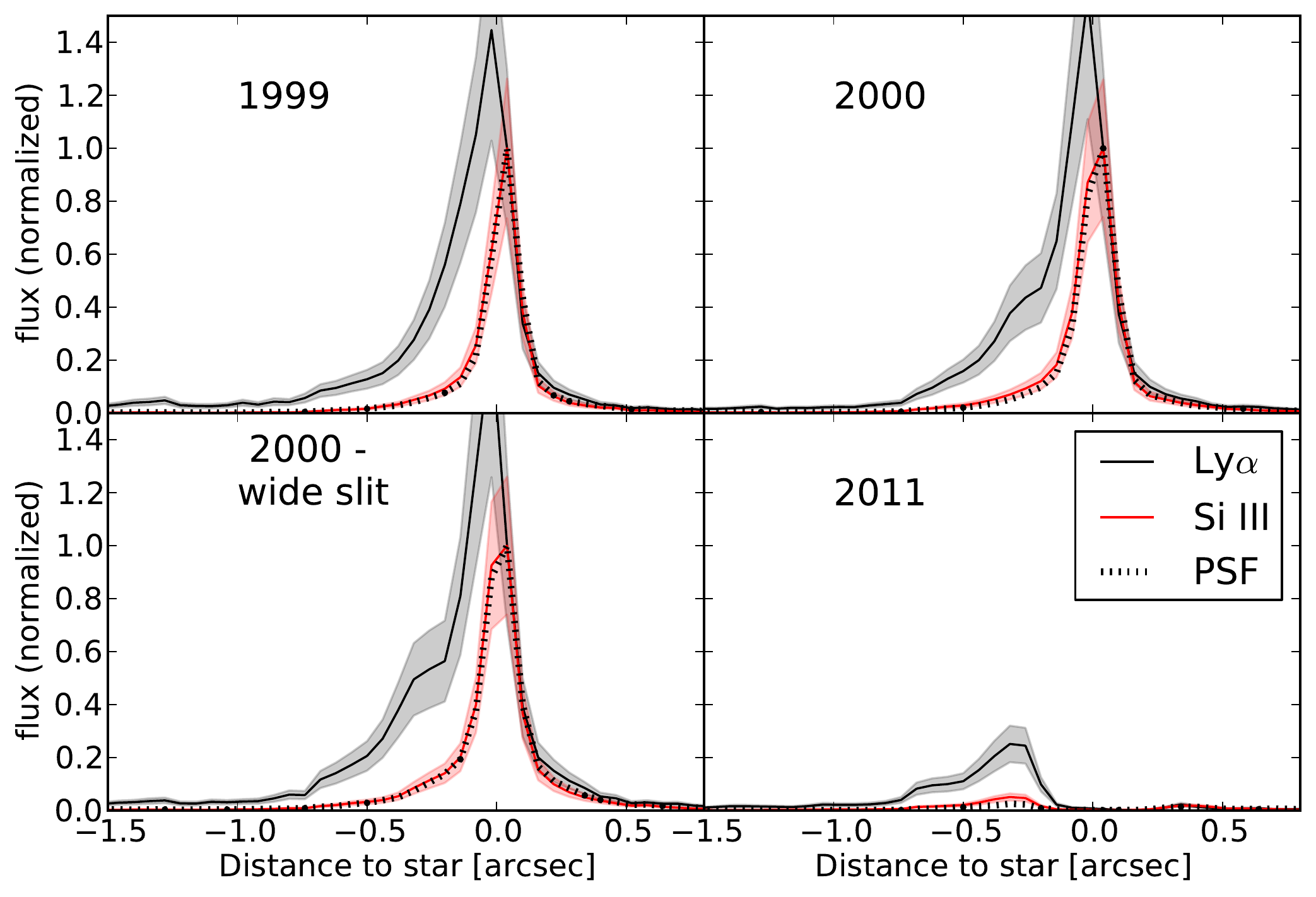}}
\caption{Spatial profile on the red (250 to 450~km~s$^{-1}$) side of the Ly$\alpha$ line, the \ion{Si}{iii} line and of a continuum region. All line profiles and continuum profiles are normalized (see text for details). Shaded regions indicate the statistical uncertainty for the lines. The uncertainty on the continuum profile is negligible because it is extracted from a much larger region. \label{fig:spatialprofile}}
\end{figure}

After subtracting the spectrum of the unresolved source \citet{2000ApJ...542L.115D} find red-shifted Ly$\alpha$ emission to the south-west of the central source in the direction of the blue-shifted jet. 
Figure~\ref{fig:spatialprofile} shows the spatial profile of the
emission on the red-shifted side of the Ly$\alpha$ line in comparison
with the continuum emission, which originates from the star. The
continuum and lines are normalized at 0.04\arcsec{}, where the continuum has the maximum of its spatial profile in our spatial binning. The red-shifted Ly$\alpha$ emission shows extended emission on the side of the approaching jet in all observations. The blue-shifted side is also extended (see Fig.~\ref{fig:PVD}), but since the jet is seen on this side we cannot distinguish if this emission is entirely due to the jet or if there is an additional component similar to that shown in Fig.~\ref{fig:spatialprofile}. In contrast, the \ion{Si}{iii} line is consistent with the continuum, which is not spatially extended.

The two observations taken in 2000 are less than a day apart and the spatial profiles have a very similar shape. Thus, it is unlikely that the difference in flux between them is due to temporal variability. Rather, it is caused by the aperture. One observation used a slit width of of 0.2\arcsec, the other 0.5\arcsec{}. Both spatial profiles are very similar, so the bulk of the emission in the direction perpendicular to the slit and thus also perpendicular to the jet, must be contained within 0.2\arcsec{}.

We searched for a similar feature in the \ion{Si}{iii} line at 1206.51~\AA{}, but the projections of line and background are fully compatible here.

\subsection{Optical emission}
\label{sect:opticalemission}
PVDs of [\ion{O}{i}] 6300\AA{}, H$\alpha$, and the [\ion{S}{ii}] lines at 6716\AA{} and 6731\AA{} are shown in Fig.~\ref{fig:PVD} (right). Stray light is strongly reduced due to the occulting wedge used in the observations, but the inner few arcseconds of the long-slit data are contaminated by Airy rings caused by refraction of the bright central star. We fit and remove the Airy rings, but the residual image shows a higher noise level at these positions.

On first sight, the H$\alpha$ emission seems extended in the direction of the blue-shifted jet -- not only for the jet velocity, but indeed for a wide range from -300 to +300~km~s$^{-1}$. A projection similar to Fig.~\ref{fig:spatialprofile} shows, that this apparent extension is indistinguishable from the spatial profile of the continuum. The only reason why it shows up prominently here is that the H$\alpha$ line is far brighter than the stellar continuum. 

Two knots can be identified in the counter-jet; no feature is seen in the jet itself (table~\ref{tab:knotpos}). The strongest knot (which we call B2) peaks at 3.3\arcsec{}, very close to the outer edge of the disk as seen in scattered light \citep{2000ApJ...544..895G}. Even without contamination by the Airy rings, the disk would block any emission from the counter-jet further in, thus we cannot say if the observed position of B2 represents the peak of the knot or if we merely observe the end of a more extended structure. The second knot (B) is located at 8.7\arcsec{}. B2 and B show a similar spacing as B and C in the old observation \citep{2006ApJ...650..985W}. 

\begin{figure}

\resizebox{\hsize}{!}{\includegraphics{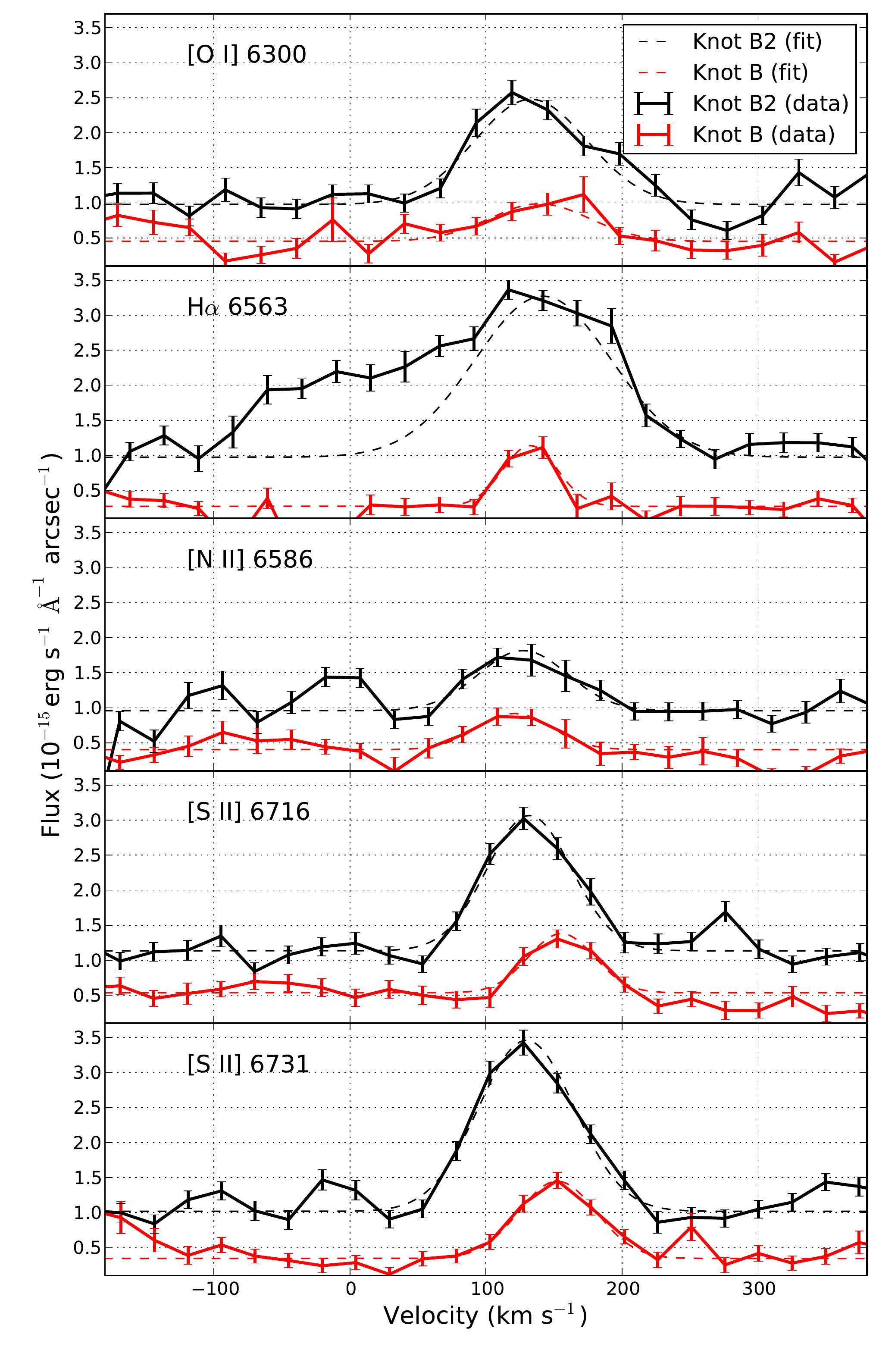}}
\caption{Spectra for H$\alpha$ and several optical FELs at the position of the two brightest knots. The main component is fitted with a Gaussian.\label{fig:optspec}}
\end{figure}

Using the extraction regions 2.6-3.7\arcsec{} and 8.1-9.0\arcsec{} we extract spectra for B2 and B, respectively (Fig.~\ref{fig:optspec}). In addition to the lines shown in the PVD we include [\ion{N}{ii}] 6586\AA{}. Gaussian fits to the lines are given in table~\ref{tab:fluxes}. H$\alpha$ is described by two Gaussians. The table reports the component with a higher velocity shift, as the inner component is most likely due to scattered stellar emission. Due to the lower spectral resolution H$\alpha$ and [\ion{N}{ii}], and the two [\ion{S}{ii}] lines at 6716 and 6731\AA{} were blended in the previous optical STIS spectra. We remeasured the line fluxes from 1999, but our values are consistent with \citet{2006ApJ...650..985W} within the errors.

The line fluxes in the table are not dereddened. The wavelength range of the optical observations is small, so a very similar correction applies to all those lines. The same $E(B - V) = 0.015$ as above leads to a flux correction of 3-4\% in the optical. This is much less than the statistical errors on the line fluxes.

\begin{table*}
\caption{\label{tab:fluxes} Fluxes in knots ($1\sigma$ uncertainties)}
\centering
\begin{tabular}{lccccc}
\hline\hline
Line & $\lambda$ & flux (1999) & flux (2011) & $v$ & $\sigma$  \\
     & \AA{} &\multicolumn{2}{c}{ $10^{-16}$~erg~s$^{-1}$~cm$^{-2}$} & km~s$^{-1}$ &km~s$^{-1}$\\
\hline
\multicolumn{6}{c}{jet: -3 to -1 \arcsec{}}\\
\hline
Ly$\alpha$ & 1216 & ...& $49_{-5}^{+2}$ & $353_{-3}^{+7}$ & $40_{-4}^{+2}$\\
\ion{Si}{iii}     & 1207 & ... & $16$\tablefootmark{a} & =353\tablefootmark{a} & =40\tablefootmark{a} \\
\hline
\multicolumn{6}{c}{jet: -5.1 to -3.3\arcsec{}}\\
\hline
Ly$\alpha$ & 1216 & ... & $32\pm2$ & $331_{-9}^{+6}$ & $77_{-2}^{+5}$\\
\ion{Si}{iii}     & 1207 & ... & $<4$\tablefootmark{a}  & =331\tablefootmark{a} & =77\tablefootmark{a}\\
\hline
\multicolumn{6}{c}{counter-jet: B2}\\
\hline
[\ion{O}{i}] & 6300 & ... &$6.3\pm 0.6$ & $130\pm10$ & $41_{-2}^{+5}$\\
H$\alpha$ & 6563    & ... &$8.4_{-0.1}^{+5.7}$ & $144_{-13}^{+3}$&$50\pm4$  \\
{}[\ion{N}{ii}] & 6586 & ... &$2.9_{-0.5}^{+0.8}$ & $127_{-10}^{+1}$ & $33_{-6}^{+13}$\\
{}[\ion{S}{ii}] & 6716 & ... & $6.4_{-0.5}^{+0.9}$  & $132\pm4$ & $31\pm3$\\
{}[\ion{S}{ii}] & 6731 & ... & $9.1\pm0.6$& $130\pm3$ & $35\pm3$\\
\hline
\multicolumn{6}{c}{counter-jet: B}\\
\hline
[\ion{O}{i}] & 6300 & $2.3_{-1.0}^{+1.1}$ & $2.0_{-0.5}^{+0.7}$ &$139_{-20}^{+3}$ &$36_{-3}^{+16}$\\

H$\alpha$ & 6563   & $21\pm2.0$\tablefootmark{b} & $2.0_{-0.3}^{+0.4}$ &$132\pm4$&$20.02_{-0.01}^{+6}$\\
{}[\ion{N}{ii}] & 6586 & $4.8_{-1.1}^{+1.5}$ &$1.4_{-0.3}^{+0.6}$ &$119\pm7$&$25.7_{-0.3}^{+21}$\\
{}[\ion{S}{ii}] & 6716 & \multirow{2}{*}{${\bigr \rbrace}20_{-2.3}^{+1.3}$} &$2.2_{-0.3}^{+0.4}$&$156\pm7$&$24\pm7$\\
{}[\ion{S}{ii}] & 6731 &   &$3.4\pm0.4$ &$153_{-5}^{+3}$&$28_{-3}^{+5}$\\
\hline
\end{tabular}
\tablefoot{
\tablefoottext{a}{The \ion{Si}{iii} lines is too weak to fit without constraints. Thus, we fix $v$ and $\sigma$ at the values found for Ly$\alpha$ at the same position.}
\tablefoottext{b}{In the \texttt{G750L} grating used in 1999, the H$\alpha$ line is blended with [\ion{N}{ii}] 6548\AA{}. This contribution is subtracted assuming the quantum mechanical ratio [\ion{N}{ii}]6584\AA{}/[\ion{N}{ii}]6548\AA{} = 3.0.}}
\end{table*}

In knot B2 in the counter-jet the velocities of all lines are roughly compatible with a systematic knot velocity of 130~km~s$^{-1}$. The forbidden lines all have a similar width, only the H$\alpha$ line is significantly wider (table~\ref{tab:fluxes}). In knot B the [\ion{S}{ii}] lines seem slightly more red-shifted than the other observed lines, however, the errors given in the table are statistical only.
In general, the velocities in the optical are smaller and narrower than in the Ly$\alpha$ line. \citet{2006ApJ...650..985W} already found a velocity asymmetry between jet and counter-jet but at much lower level and with larger errors ($223\pm81$ vs. $171\pm89$~km~s$^{-1}$). Apparently, the Ly$\alpha$ emission in the jet traces a faster component than the FELs. In our observation the FUV features in the jet are not detected in the optical, thus a comparison of FUV and optical lines in the same knot is not possible.
Velocity asymmetries are not unusual. The ratio of jet velocity and
  counter-jet velocity is $>1.5$ for about half of all CTTS with
  optically detected outflows. Those differences seem to be intrinsic
  to the launching process and cannot be explained by interactions
  with the circumstellar matter close to the central source \citep{1994ApJ...427L..99H}.

\subsubsection{Proper motion and velocity of knot B}
\label{sect:knotBproper}
Knot B was seen at 4.3\arcsec{} in 1999. For reasonable values
of its proper motion it should be located a few arcsec further out in
the new STIS observation. Indeed, a knot is located at 8.7\arcsec{}
and we identify this as knot B with a proper motion of
$0.37\pm0.01$\arcsec{}~yr$^{-1}$ which agrees well with the proper motion of knot C of $(0.34\pm0.02)$\arcsec{}~yr$^{-1}$. Thus, it seems that knots in the jet move consistently with similar velocities.

We take the velocity along the line of sight as the mean of all observed line shifts for knot B (table~\ref{tab:fluxes}) and derive a total space motion of 250~km~s$^{-1}$ for knot B and an inclination of 56\degr. The proper motion measures the motion of the shock front, while the line shift shows the motion of the gas. Thus, the above calculation requires that the shock front is co-moving with the material. As shown below, the relative shock speed is low and thus this approximation is valid.
The measured values agree with previous measurements of the jet speed and inclination.

\subsubsection{Physical conditions in knot B}
While the observed surface luminosity in knot B2 is about 2-3 times
higher than in knot B, the line ratios are very similar, which
indicates a similar density and temperature, although knot B is nearly
twice as far from the central source as knot B2 (integrated
  fluxes are given in table~\ref{tab:fluxes}). A decade ago, when
knot B was located close to the current position of knot B2, its
surface luminosity in H$\alpha$ and [\ion{S}{ii}] was significantly
higher compared with knot B today. The surface luminosity of knot B in
the [\ion{S}{ii}] and [\ion{O}{i}] lines reduced by a factor of 2, in
[\ion{N}{ii}] by a factor of 4, but in H$\alpha$ by an order of magnitude between 1999 and 2011. 

\subsubsection{Electron densities}
\label{sect:ne}
The ratio of the [\ion{S}{ii}] lines at 6716\AA{} and 6731\AA{} is density sensitive with only a very modest temperature dependence. The differential reddening between those two lines is negligible, because HH~409 is only weakly reddened and the wavelength is very similar. Unfortunately, these lines are unresolved in the early STIS observations, so that we do not have a comparison value for knot B from an earlier epoch. However, in 2004 the density of knot C was measured from ground-based data to $n_e = 1400\pm400$~cm$^{-3}$ \citep{2006ApJ...650..985W}. With the observed line values from table~\ref{tab:fluxes} we find $n_e = 2200_{-600}^{+1000}$~cm$^{-3}$ and $n_e = 3000_{-1000}^{+3000}$~cm$^{-3}$, for knots B2 and B, respectively, using the CHIANTI database \citep{CHIANTI,2012ApJ...744...99L}. Below we calculate the total density $n$ from $n_e$ and the ionization fraction.

\subsubsection{Shock velocities}
We estimate the shock velocities and the ionization fraction in the knots using the models of \citet{1994ApJ...436..125H}. They contain updates for some optical lines over the simulations by \citet{1987ApJ...316..323H}, which we used to interpret the Ly$\alpha$/\ion{Si}{iii} ratio. \citet{1994ApJ...436..125H} present a series of plots for different line ratios that provide a diagnostic of the ionization fraction and the shock velocity. The models are calculated for three different pre-shock densities and for different magnetic fields. Given the estimate of the electron density above, we use the models for a pre-shock density of $10^3$~cm$^{-3}$. The magnetic field in the shock zone is degenerate with the shock speed because a field parallel to the shock front inhibits strong compression. We assume a weak field ($<0.1$~mG) here.

\begin{table}
\caption{\label{tab:shock} Ionization fraction and $v_{shock}$}
\centering
\begin{tabular}{lrrr}
\hline\hline
Line ratio & B (1999) & B (2011) & B2(2011) \\
\hline
\multicolumn{4}{c}{Ionization fraction}\\
\hline
[\ion{O}{i}]/H$\alpha$ & 0.2 & 0.02 & 0.03\\
{}[\ion{S}{ii}](6716+6731)/H$\alpha$ & 0.07 & 0.02 & 0.04\\
{}[\ion{N}{ii}]/[\ion{O}{i}] & 0.3 & 0.1 & 0.1\\
\hline
\multicolumn{4}{c}{shock velocity [km~s$^{-1}$]}\\
\hline
[\ion{O}{i}]/H$\alpha$ & 50 & 30 & 40\\
{}[\ion{S}{ii}](6716+6731)/H$\alpha$ & 40 & 30 & 30\\
{}[\ion{N}{ii}]/[\ion{O}{i}] & 70 & 30 & 30\\
{}[\ion{S}{ii}]6716/[\ion{S}{ii}]6731 & ... & 45 & 40\\
\hline
\end{tabular}
\end{table}

Table~\ref{tab:shock} shows the ionization fractions $<I>$ (averaged over the emitting area) and the shock velocities $v_{shock}$ from the \citet{1994ApJ...436..125H} models. The shock velocities derived from different line ratios agree well. Knot B and B2 in the most recent observation have a very similar shock speed around 30~km~s$^{-1}$, less than the shock speed of knot B in 1999. The differences between the ionization fractions are much larger. Again, the knot B in 1999 was more highly ionized than any knot in 2011, however, the recent values span a range of a factor of five. This could be due to an underestimate of the H$\alpha$ flux in the fitting process. In knot B2 the fitted Gaussian has a larger velocity than all other lines. Thus, a larger fraction of the flux is described by the Gaussian of a stationary component, which we interpret as scattered light from the central source. On the other hand, the relative errors on the line fluxes for weak lines are large and the differences in $<I>$ in table~\ref{tab:shock} are only marginally significant. In the following we adopt a shock velocity of 50~km~s$^{-1}$ for knot B in 1999 and 35~km~s$^{-1}$ for knot B and B2 in 2011. The shock models predict a compression $<C>$ of gas (averaged over the [\ion{S}{ii}] emitting area) of a factor 50 and 25 for these shock velocities.

\subsubsection{Mass loss rates}
There are several different methods to estimate the mass flux in a jet. In this section we follow \citet{1994ApJ...436..125H}. First, the mass flux can be estimated from the total flux in a line, e.g. [\ion{O}{i}]. We calculate the total luminosity and assume that the plasma is emitting close to the peak formation temperature of this line and that the total oxygen abundance is close to solar. This calculation  misses flux components that are located outside the aperture, ionized, too cool to emit [\ion{O}{i}] 6300\AA{} or have densities above the critical density for [\ion{O}{i}], e.g. in a dense and fast component very close to the jet axis. While we know little about temperature components outside of the formation range of FELs, the density is most likely below the critical density ($10^6$~cm$^{-3}$) given the estimates above. We use
\begin{eqnarray}
\dot M  & = & 5.95\times10^{-8} \left(\frac{n_e}{10^3\textnormal{ cm}^{-3}}\right)^{-1}\left(\frac{L_{6300}}{10^{-4} L_{\sun}}\right) \nonumber\\
 & & \times \left(\frac{v_{sky}}{100\textnormal{ km s}^{-1}}\right)\left(\frac{l_{sky}}{10^{16}\textnormal{ cm}}\right)^{-1} M_{\sun} \textnormal{ yr}^{-1} \ ,
\end{eqnarray}
which is eqn. 10 from \citet{1994ApJ...436..125H}. Alternatively, a very similar calculation can be done using the [\ion{S}{ii}] lines. If they are in the high-density limit ($>10^4$~cm$^{-3}$), then the formula does not depend on the density in the knot \citep[][eqn A.10]{1995ApJ...452..736H}: 
\begin{eqnarray}
\dot M  & = & 4.5\times10^{-9} \left(\frac{L_{6731}}{10^{-4} L_{\sun}}\right)\left(\frac{v_{sky}}{100\textnormal{ km s}^{-1}}\right) \nonumber\\
 & & \times \left(\frac{l_{sky}}{10^{16}\textnormal{ cm}}\right)^{-1} M_{\sun} \textnormal{ yr}^{-1} \; .
\end{eqnarray}

Finally, we can calculate the mass flux through an area $A$, when the gas moves at a velocity $v$:
\begin{equation}
\dot M = \rho v A = \mu m_H n v A \ ,
\end{equation}
where $\rho= \mu m_H n$ is the mass density, calculated from the mean molecular weight $\mu = 1.24$, the mass of the hydrogen atom $m_H$ and the atom/ion number density $n$. The difficulty here is to estimate $n$ from the measured electron density $n_e$. This can be done using ionization $<I>$ and the compression $<C>$ from the shock models of  \citet{1994ApJ...436..125H}. These authors recommend to scale the density with $\sqrt{<C>}$ because this is more appropriate for a clumpy medium than scaling with the compression factor itself. We get:
\begin{equation}
n = \frac{n_e}{\sqrt{<C>}<I>}\;.
\end{equation}

Since we do not measure the size of the jet perpendicular to the direction of the flow, we use a radius of 0.15\arcsec{}, which is the width of the innermost knots resolved by \citet{2006ApJ...650..985W}. We take the total space motion of knot B as the flow speed of material. Table~\ref{tab:mdot} gives the mass flux in the knots B and B2 estimated from all three methods. 
Given the simplifications the difference in the mass flux measured is not surprising. The densities of knot B and B2 are very similar, but their luminosities differ. This indicates, that we see a larger emitting volume in knot B2. In knot B the mass flux calculated from the flow is much larger than the mass flux from the line luminosities. This could indicate, that either a larger part of knot is too cool to radiate in [\ion{O}{i}] or [\ion{S}{ii}] or that the knot has a small filling factor, i.e. part of the volume contains cool material unrelated to the jet \citep{2006A&A...456..189P}.

\begin{table}
\caption{\label{tab:mdot} Mass flux in the knots}
\centering
\begin{tabular}{lrrr}
\hline\hline
 & B (1999) & B (2011) & B2(2011) \\
\hline
$n_e$ [cm$^{-3}$] & $\approx 3000$ & 3000 & 2200 \\
$<C>$ & 50 & 25 & 25 \\
$<I>$ & 0.2 & 0.05 & 0.05\\
$l_{sky}$ [$10^{14}$ cm] & 6 & 6 & 6\\
$v_{sky}$ [km s$^{-1}$]& 190 & 190 & 210\\
$L_{6300}$ [$10^{-7}L_{\sun}$] & 2 & 1 & 3 \\
$L_{6731}$ [$10^{-7}L_{\sun}$] & ... & 2 & 5 \\
$n$ [cm$^{-3}$] & ... & $12\times10^3$ & $9\times10^3$ \\
$\dot M_{6300}$ [$10^{-9}M_{\sun}$~yr$^{-1}$] & 1.3 & 0.6 & 2.8\\
$\dot M_{6731}$ [$10^{-9}M_{\sun}$~yr$^{-1}$] & ... & 0.3 & 0.8\\
$\dot M_{flow}$ [$10^{-9}M_{\sun}$~yr$^{-1}$]& ... & 2.8 & 2.2\\
\hline
\end{tabular}
\end{table}

\section{Discussion}
\label{sect:discussion}

\subsection{The central star}
HD 163296 is remarkably constant over nearly 200~ks of X-ray observing time. This is consistent with the abundance pattern of an inactive star and the cool temperatures. In contrast, CTTS typically are active stars with frequent flaring and an inverse FIP abundance pattern. A jet collimation shock could explain some of the soft emission. 
However, no offset of the X-ray emission from the central star is seen and we estimate an upper limit of 30~AU. This is stricter than the number given in \citet{HD163296} and we can thus calculate a more stringent limit on the conditions at the jet base, assuming that a shock at the jet base is the origin of the central soft X-ray component. The post-shock cooling length $d_{\mathrm{cool}}$ depends only on the pre-shock density $n_0$ and shock temperature and can be written as \citep{2002ApJ...576L.149R}:
\begin{equation}
d_{\mathrm{cool}} \approx 8 \mathrm{ AU} 
    \left(\frac{10^5\mathrm{ cm}^{-3}}{n_0}\right) 
    \left(\frac{v_{\mathrm{shock}}}{400\textnormal{ km s}^{-1}}\right)^{4.5} \; .
\end{equation}
\citet{HD163296} derive $v_{\mathrm{shock}} = 400$~km~s$^{-1}$, thus the lower limit on the density is $n_0 > 2\times 10^4$~cm$^{-3}$. The flow time through this post-shock region would be about a year. In turn, the density sets an upper limit on the radius $R$ of the shock front assuming a circular shock at the base of a cylinder \citep[for details see][]{dgtau}
\begin{equation}
R \approx 1.6 
        \left(\frac{10^5 \mathrm{cm}^{-3}}{n_0}\right)^{0.5}
        \left(\frac{VEM}{10^{52} \textnormal{cm}^{-3}}\right)^{0.5} 
        \left(\frac{0.21 \mathrm{keV}}{\mathrm{k}T}\right)^{1.125} \;,
\label{eqnr}
\end{equation}
where $R$ is given in AU. With the VEM from table~\ref{tab:3vapecem}
this leads to an upper limit $R<5$~AU. This radius would require a
rather large opening angle at the base of the jet, thus it is likely
that the actual densities are higher than the lower limit derived
here. The mass flux required to power the X-ray emission is
  $6\times10^{-11}M_{\sun}$~yr$^{-1}$, only about a tenth of the mass
  flux in the outer optical knots (table~\ref{tab:mdot}). Thus, this
  seems to be powered by a different (faster and less massive) component of the outflow than the
optical knots in the jet.

\subsection{The one sided Ly$\alpha$ emission}
We find a significant extension of the Ly$\alpha$ line in the direction of the approaching jet in all observations. The signal on the blue-shifted side of the emission line could partly be caused by the jet itself, but the red-shifted emission has to come from a different origin. The emission is seen between 0.1\arcsec{} and 1\arcsec{}, which corresponds to 12 and 120~AU if the emitting structure is in the plane of the sky. Perpendicular to the jet the emission is mostly contained in the inner 0.2\arcsec{}. The emission reaches as far out as 400~km~s$^{-1}$ as seen e.g.\ at the position of the first knot (see Fig.~\ref{fig:FUVspec}, where this feature is labeled ``scattered emission?'') but any signal within 150~km~s$^{-1}$ of the line center is hidden behind interstellar absorption and geocoronal emission.

This feature was noticed already by \citet{2000ApJ...542L.115D} who suggest it to be caused by infall or a poorly collimated wind. We disagree with this explanation because of the large width of the observed line. Thermal broadening cannot produce the width since hydrogen would be fully ionized at the required temperatures and pressure broadening only becomes relevant for densities of the order $10^{13}$~cm$^{-3}$ \citep{1973ApJS...25...37V}. Thus, the line width must be explained by kinematic motions but at distances of a few AU where the feature is seen the gravity of the central star provides an accretion flow speed of a few km~s$^{-1}$ only, much less than the width of the emission line.

The disk obscures the far side of the system, so that we only see the approaching wind. If we were looking pole-on all outflows on the visible side would be blue-shifted, but since the inclination is 50-60\degr{} some outflows might appear red-shifted. In the extreme case of a flat disk and a disk-wind moving in parallel to the disk the flow speed still has to reach 500~km~s$^{-1}$ in order to achieve a line-of-sight velocity of  400~km~s$^{-1}$. 
Not only is the limiting case of a flat disk unrealistic, the required velocity for this outflow is also much larger than the jet velocity. Although we do not understand the jet collimation process in detail, current models suggest that the collimation happens within a few stellar radii \citep[e.g.][]{2012MNRAS.420.2020L} and commonly the less-collimated layers of the outflow are much slower. One mechanism for such an outflow is a photo-evaporative disk wind, where the upper layer of the disk is heated by the stellar radiation. This layer is hot enough to emit in Ly$\alpha$, but predicted velocities for CTTS are only a few km~s$^{-1}$ \citep{2010MNRAS.406.1553E,2011ApJ...736...13P}. Even for HAeBe, which are much brighter in the UV, these winds will not reach the required velocities. Additionally, all the above models would predict an equally strong blue-shifted emission which is not observed. 

These difficulties make it promising to propose a model, where the Ly$\alpha$ flux seen at some distance to the central source does not originate there but is scattered emission from the central source, since a spectrum extracted at the stellar position shows that the Ly$\alpha$ line is wide enough. Presumably, most of the hydrogen emission originates in the accretion funnel and the accretion shock \citep{1998ApJ...492..743M}. However, dust scattering is not wavelength selective and would thus lead to comparable spatial profiles for the FUV continuum and the wings of the Ly$\alpha$ line in contrast to Fig.~\ref{fig:spatialprofile}. (The same argument makes an instrumental artifact unlikely.) 
This can be overcome, if the Ly$\alpha$ line is formed in the accretion funnels above the plane of the disk. The stellar continuum would be absorbed by the inner disk wall, while the accretion funnels could reach above the rim and illuminate the disk surface out to the required radius. \citet{2008ApJ...682..548W} observe dark lanes in scattered optical light a few arcseconds from the central source, which correlate with the overall disk luminosity and are consistent with shadowing by a time-variable inner disk wall. An alternative explanation is a dusty disk wind, with dust grains entrained in the outflow \citep{2012ApJ...758..100B}. A flared disk or a dusty disk wind would intercept a large fraction of the total Ly$\alpha$ flux. In principle, we should see Ly$\alpha$ from the entire disk surface but it is only observed in the negative spatial direction (below the star in Fig.~\ref{fig:PVD} and to the left in Fig.~\ref{fig:spatialprofile}). This is the far side of the disk that we see through the jet, so the light needs to be back-scattered to reach us. However, in the UV dust is strongly forward scattering \citep{2003ApJ...598.1017D} and thus the strongest signal should be observed in the positive spatial direction. The scattering angles in a dusty wind are slightly more favorable than in a flared disk.

Alternatively, scattering by dust in the jet would explain naturally why we see the extended emission on the side of the jet only and also why it seems not to be extended perpendicular to the jet axis. Also, the required scattering angle is the inclination of the jet (about 50\degr{}), which is in the preferred range of angles. 
A potential problem with this scenario is that there is no direct observational evidence for dust in the jet. Theoretically, if the jet material is launched
close enough to the central star, all dust may be destroyed near the jet base, by stellar irradiation and/or thermal evaporation, although detailed calculations are needed to quantify the dust content of the jet.

Since the spatial profiles from observations between 1999 and 2011, 12 years apart, are compatible it is unlikely that a feature in the disk, e.g. a density enhancement on the far side of the disk, causes the scattering to be much more efficient on one side than on the other.

We conclude that several components of dust scattering, wind emission and wind absorption with different speeds and launching radii need to be present at the same time to produce the observed signature.

\subsection{Time evolution of the knots}
Knot B is the only knot observed in 1999 and in 2011. In section~\ref{sect:knotBproper} we already derived its proper motion. In this time period the observed surface luminosity decreased significantly. This could be due to intrinsic dimming or indicate that the jet width increased beyond the width of the slit used in the observations. The lowest difference --a factor 2-- is observed for [\ion{O}{i}], the largest difference --a factor 10-- is observed for H$\alpha$. Consequently, the ionization fraction and shock speed deduced from the line ratios are lower in 2011 than in 1999. 

Except for the [\ion{S}{ii}] doublet no element is observed with more than one line, thus the temperature of the emitting material cannot be measured independent of the abundances. However, since all observed elements are neutral or singly ionized, plasma with temperatures between 5000 and 25000~K must be present. In this energy range, atomic lines are the dominant contributors to line cooling. As an estimate, we use the radiative loss rate $\Lambda = 5\times 10^{-23}$~erg~cm$^3$~s$^{-1}$ which is valid for this temperature range within a factor of a few. The cooling time $\tau$ can be calculated as the ratio of the thermal energy content of the gas and the radiative loss as
\begin{equation}
\tau = \frac{3/2 n\textnormal{k}T}{\Lambda n n_e} \approx 0.4 \textnormal{ yr ,}
\end{equation}
where k is the Boltzmann constant and we use the $n_e$ and $n$ estimated in Sect~\ref{sect:ne}. The short time scale shows that the knot must be continuously reheated by a shock that converts the kinetic energy of the jet into thermal energy. Ultimately, the radiation is thus powered from the kinetic energy of the jet. The kinetic energy density of the material is 300 times larger than the thermal energy density. Indeed, $v_{sky}$ and the shock velocity are both lower in the 2011. This is consistent with a jet that decelerates as its kinetic energy is radiated away. An additional consequence is that the measured proper motion, which is determined from the position of the shock front, is not the same as the proper motion of the mass in the jet. 

The knots A and C are not visible in the new observation. As shown above the cooling times for gas in the knots are very short, so no detection is expected, if the shock fronts in these knots are no longer strong enough to power the radiation.

Due to the low signal in the X-ray observations we cannot derive meaningful limits on the density and cooling time, but we note that in DG~Tau the outer X-ray knots are less luminous than the inner ones \citep{2008A&A...478..797G} and a cooling of the X-ray plasma as it travels outward has been observed by \citet{2011A&A...530A.123S}.

\subsection{Long term evolution of the jet}
In ground-based observations \citet{2006ApJ...650..985W} are able to identify fainter knots further away from the driving source and they estimate a dynamical age of HH~409 of at least 80~yr. Our observations are not as sensitive to fainter structures, but in the inner jet we can confirm that the general shape of the jet has not changed over one decade. We have discovered a new knot (B2), which has a similar velocity as the previously innermost knot B. The distance to the preceding knot is also comparable, indicating a periodicity for knot launching. However, knot B2 has a lower shock velocity and compression factor, but slightly larger electron density and mass loss rate than knot B had at a comparable distance from the driving source. This shows that the launching conditions must have changed slightly.

\subsection{Implications for jet launching}
Most theoretical models rely on magnetic fields to explain the jet launching and collimation. Yet, it is unclear if the central star HD~163296 posses a large scale magnetic field. On the one hand, none was found in the observations of \citet{2007A&A...463.1039H}, on the other hand, we observe a constant X-ray flux with temperatures above 2~keV, which is beyond the heating from an accretion shock and is presumably caused by magnetic reconnection and thus indicates the consistent presence of stellar magnetic fields.

We know that knot B2 was launched within a few years of the optical polarimitry and the X-ray observations: If it has moved with constant proper motion, then it was launched in 2002, while the observations of \citet{2007A&A...463.1039H} were performed in 2004 and we analyzed X-ray data from 2003, 2007 and 2011. If the limit on the magnetic field is representative then a strong stellar magnetic field is not a necessary pre-requisite of jet launching. Instead, \citet{2007A&A...463.1039H} report the detection of Stokes components in the Ca H and K lines, which are much wider than expected from the photosphere. If these lines are of circumstellar origin they probe the jet launching and collimation region more directly. These fields could be generated in e.g. a disk dynamo \citep{1997ApJ...475..263V}.

Alternatively, we can take the presence of hard X-rays as proof of stellar magnetic fields. The optical polarimitry concentrated on hydrogen lines, which -- at least in CTTS -- have multiple contributions from the inner edge of the disk, the accretion funnel and the star. The stellar contribution of the flux might be too small to produce a measurable Stokes signal. Thus, we argue that the jet launching and collimation in HAeBes is comparable with the process in CTTS.

\section{Summary}
\label{sect:summary}
We analyze a series of \emph{Chandra} and \emph{HST} observations of
HH~409, the jet from the Herbig Ae star HD~163296. In X-rays the star
itself is very stable and the observed temperatures and emission
measures are fully compatible with previous observations. No extension
beyond the PSF is found for the central star, which limits any
unresolved sources with a significant fraction of the total luminosity
to a maximum separation of 30~AU. This is important, because an
\emph{XMM-Newton} spectrum showed that the soft X-ray emission
originates above the stellar surface, possibly in the jet collimation
region. While in 2003 a weak feature on the edge of the stellar PSF is
observed at a distance of -4\arcsec{}, it is unclear if emission at
that position is present in 2011. \citet{2005ApJ...628..811S} claim
the detection of a knot in the jet at -7.2\arcsec{}; our analysis
confirms this. There is a tentative grouping of three photons at
-10.6\arcsec{} in 2011. This position is consistent with the 2003
knot, if it propagates with the same proper motion as the optical knots. In any case, the luminosity of the knot dropped significantly since 2003, indicating a short cooling time.

In the UV the Ly$\alpha$ line is the most prominent feature. In the
new observations we detect two knots in the jet, but not the
counter-jet. In addition, there is extended emission on the side of
the approaching jet that is red-shifted up to 400~km~s$^{-1}$ and can
be seen out to about 1\arcsec{}. On the blue-shifted side the emission
is much weaker. Neither red nor blue-shifted emission is seen in the direction of the counter-jet. It requires multiple components of accretion, a wide-angle wind or dust scattering of the Ly$\alpha$ line from the accretion spot to explain this feature.

In the optical we detect H$\alpha$ and several forbidden emission lines. We detect a new knot, called B2, and re-observe the previously detected knot B. The distance between those two knots is similar to the distance between knot B and the preceding knot. This nicely fits into a series of knots in the counter-jet with almost equal spacing which were launched over several decades \citep{2006ApJ...650..985W}. Knot B moves with a proper motion compatible to the previous values for knot C. Knot B is significantly fainter now than in 1999, particularly in H$\alpha$. The line ratios indicate a lower temperature. A significant spatial expansion, which would bring the size of the knot beyond the width of the slit, could also contribute to the fading. Still, the cooling time scale is less than a year, so knot B needs to be continuously re-heated. This supports a picture where the knots are shock fronts traveling though a jet of continuously launched, cool material. Over time, these shocks radiate away their energy, so the post-shock cooling zone formed behind them has consecutively lower temperatures.

Knot B2 today is seen at almost the same distance from the central star as knot B was a decade ago. Yet, knot B2 has a lower ionization and radial velocity, but a larger mass flux than knot B had a decade ago. Thus, the jet launching conditions must have changed in this period.

\begin{acknowledgements}
Support for this work was provided by the National Aeronautics and Space Administration (NASA) through Chandra Awards Number GO2-13015X and TM0-11002X issued by the Chandra X-ray Observatory Center, which is operated by the Smithsonian Astrophysical Observatory for and on behalf of NASA under contract NAS8-03060. Further support comes from grant GO-12186.01-A from the Space Telescope Science Institute, which is operated by the Association of Universities for Research in Astronomy, Inc., under NASA contract NAS5-26555. PCS acknowledges support from the DLR under 50~OR~1112. Some of the data presented in this paper were obtained from the Mikulski Archive for Space Telescopes (MAST).
\end{acknowledgements}


\bibliographystyle{aa}
\bibliography{../articles}

\end{document}